\documentclass[useAMS,usenatbib]{mnras}
\usepackage{graphicx}
\usepackage{color}

\newcommand{\ml}{$\rm M/L$}
\newcommand{\mlr}{$\rm M_\star/L_r$}

\newcommand{\simlt}{\lower.5ex\hbox{$\; \buildrel < \over \sim \;$}}
\newcommand{\simgt}{\lower.5ex\hbox{$\; \buildrel > \over \sim \;$}}

\newcommand{\mgf}{$\rm Mg4780$}
\newcommand{\atio}{$\rm aTiO$}
\newcommand{\tioi}{$\rm TiO1$}
\newcommand{\tioiio}{$\rm TiO2_{SDSS}$}
\newcommand{\tioii}{$\rm TiO2$}

\newcommand{\feh}{$\rm FeH0.99$}

\newcommand{\nai}{$\rm NaD$}
\newcommand{\nad}{$\rm NaD$}
\newcommand{\naii}{$\rm NaI8190$}
\newcommand{\naiii}{$\rm NaI1.14$}
\newcommand{\naiv}{$\rm NaI2.21$}

\newcommand{\mgfep}{$\rm [MgFe]'$}
\newcommand{\mgfe}{$\rm [Mg/Fe]$}

\newcommand{\nafe}{$\rm [Na/Fe]$}
\newcommand{\tife}{$\rm [Ti/Fe]$}

\newcommand{\cfe}{$\rm [C/Fe]$}

\newcommand{\hbo}{$\rm H\beta_o$}
\newcommand{\hb}{$\rm H\beta$}
\newcommand{\hgamma}{$\rm H\gamma$}

\newcommand{\hgf}{$\rm H\gamma_F$}

\newcommand{\kms}{\,km\,s$^{-1}$}
\newcommand{\afe}{$[\rm \alpha/{\rm Fe}]$}
\newcommand{\afep}{$\rm [Z_{Mg}/Z_{Fe}]$}

\newcommand{\gammab}{$\rm \Gamma_b$}

\newcommand{\zh}{$\rm [Z/H]$}

\newcommand{\res}{$\rm R_{\rm e,S}$}
\newcommand{\reb}{$\rm R_{\rm e,B}$}
\newcommand{\ret}{$\rm R_{\rm e,T}$}
\newcommand{\mts}{$\rm m_{\rm T,S}$}
\newcommand{\mtbd}{$\rm m_{\rm T,B+D}$}
\newcommand{\ns}{$\rm n_{\rm S}$}
\newcommand{\nb}{$\rm n_{\rm B}$}
\newcommand{\ba}{$\rm b/a$}
\newcommand{\bt}{$\rm B/(B \! + \! D)$}
\newcommand{\alfa}{$\alpha$}
\newcommand{\alj}{$\rm \alpha_{Na_j}$}
\newcommand{\Teff}{$\rm T_{eff}$}

\usepackage[T1]{fontenc}
\usepackage{ae,aecompl}

\usepackage{graphicx}	
\usepackage{amsmath}	
\usepackage{amssymb}	

\title[IMF gradients in massive ETGs]{IMF radial gradients in most massive early-type galaxies.}

\author[F. La Barbera et al.]
{F. La Barbera$^{1}$\thanks{E-mail:  francesco.labarbera@inaf.it (FLB)},
A., Vazdekis$^{2,3}$, I. Ferreras$^{4}$, A. Pasquali$^{5}$,
C. Allende Prieto$^{2,3}$, \and 
I. Mart\'in-Navarro$^{6}$, D.~S. Aguado$^{7}$, 
R.~R. de Carvalho$^{8,9}$, S. Rembold$^{10}$,  \and
J. Falc\'on-Barroso$^{2,3}$, G. van de Ven$^{11}$
\\
$^{1}$INAF-Osservatorio Astronomico di Capodimonte, sal. Moiariello
16, Napoli, 80131, Italy\\
$^{2}$Instituto de Astrof\'\i sica de Canarias, Calle V\'\i a L\'actea s/n, E-38205
  La Laguna, Tenerife, Spain\\
$^{3}$Departamento de Astrof\'\i sica, Universidad de La Laguna (ULL), E-38206  La Laguna, Tenerife, Spain\\
$^{4}$Mullard Space Science Laboratory, University College London, Holmbury St Mary,
  Dorking, Surrey RH5 6NT, UK\\
$^{5}$Astronomisches Rechen-Institut, Zentrum f\"ur Astronomie, Universit\"at Heidelberg, M\"onchhofstr. 12-14, D-69120 Heidelberg, Germany\\
$^{6}$ University of California Observatories, 1156 High Street, CA-95064, Santa Cruz, USA\\
$^{7}$ Institute of Astronomy, University of Cambridge, Madingley Road, CB3 0HA Cambridge, UK\\
$^{8}$ NAT - Universidade Cruzeiro do Sul / Universidade Cidade de Sa\~o Paulo\\
$^{9}$ Instituto Nacional de Pesquisas Espaciais (INPE)\\
$^{10}$ Departamento de F \'isica, CCNE, Universidade Federal de Santa Maria, 97105-900, Santa Maria, RS, Brazil\\
$^{11}$ Department of Astrophysics, University of Vienna, T\"urkenschanzstrasse 17, 1180 Vienna, Austria\\
  }

\date{}

\pubyear{2019}

\begin{document}
\label{firstpage}
\pagerange{\pageref{firstpage}--\pageref{lastpage}}
\maketitle

\begin{abstract}
Using new long-slit spectroscopy obtained { with X-Shooter at ESO-VLT}, 
we study, for the first time, radial gradients of optical and Near-Infrared IMF-sensitive features 
in a representative sample of galaxies at the very high-mass end of the galaxy population. The sample consists of { seven early-type} galaxies (ETGs) at z$\sim 0.05$, with central velocity dispersion in the range $300 \lesssim \sigma \lesssim 350$~\kms . Using state-of-art stellar population synthesis models, we fit a number of spectral indices, from different chemical species { (including TiO's and Na indices)}, to constrain { the IMF slope (i.e. the fraction of low-mass stars)}, as a function of galactocentric distance, over  a radial range out to $\sim 4$~kpc.
{ ETGs in our sample show} a significant correlation of IMF slope and surface mass density. The bottom-heavy population (i.e. an excess of low-mass stars in the IMF) is confined to central galaxy regions with surface mass density above $\rm \sim 10^{10} M_\odot \, kpc^{-2}$, or, alternatively, within a characteristic radius of $\sim 2$~kpc. { Radial distance, in physical units, and surface mass density, are} the best correlators to IMF variations, with respect to other dynamical (e.g. velocity dispersion) and stellar population (e.g. metallicity) properties. 
 { Our results for the most massive galaxies suggest that there is no single parameter} that fully explains variations in the stellar IMF, but IMF radial profiles at $z \sim 0$ rather result from the complex formation and mass accretion history of galaxy inner and outer regions.
\end{abstract}

\begin{keywords}
galaxies: stellar content -- galaxies: fundamental parameters -- galaxies: formation -- galaxies: elliptical and lenticular, cD
\end{keywords}



\section{Introduction}

Our current understanding of galaxy formation and evolution rests on a
framework based on a $\Lambda$CDM expanding Universe, where the mass
budget is dominated by dark matter, but where the direct observables
originate from baryonic material, most notably stars. The creation of
stars from gas proceeds through a complex set of physical
mechanisms. One of the fundamental pieces of the star formation puzzle is
the distribution of stellar masses at birth, i.e. the initial mass
function (IMF). From basic principles, it is possible to construct
arguments that can explain the overall shape of the IMF and its
potential variations, via either  analytic models
\citep{Hopkins:13,Chabrier:14}, or more detailed hydrodynamical
simulations \citep{PadNor:02,BB:05,Krumholz:16}. However, it is
through observations that we can discriminate against the different
scenarios laid before us by the theoretical work. Observational
constraints of the IMF have been sought for a long time
\citep[see][for an overview]{Bastian:10}, including
both resolved \citep{Geha:13} and unresolved populations \citep[see,
  e.g.,][]{FF:80,Cenarro:2003}.

The traditionally adopted universality of the IMF has been challenged
both on star-forming and quiescent systems \citep[see][for a recent
 review]{AH:18}.  Regarding the latter, improved technology and
population synthesis modeling have allowed us to explore a
decades-long idea of using spectral line strengths that are sensitive
to the giant vs dwarf stellar ratio as a discriminant of the IMF in
passive populations \citep{SpinTa:71}.  This is especially relevant, as massive,
quiescent galaxies reveal a rather extreme formation history typically
described by an early, intense and short-lived formation process
\citep[see, e.g.,][]{IGdR:11}. Such formation implies substantially different physical
properties in the interstellar medium, that could, perhaps, lead to a
qualitatively different mode of star formation in these systems.  The
finding of a bottom-heavy IMF in massive early-type galaxies
\citep[hereafter ETGs,][]{Cenarro:2003,vdC:10}, later confirmed with independent
data and analyses~(see, e.g., \citealt[hereafter F13]{Ferr:13}; \citealt[hereafter LB13]{LB:13}; \citealt{Spiniello:2014})
was consistent with this scenario. These results were supported by
dynamical constraints, produced at around the same time, and based on
kinematic studies of integral field unit data of nearby ETGs
\citep{Capp:12,Capp:13}, that revealed high values of the stellar M/L
with respect to the expectations from a standard, Milky Way
IMF (see also~\citealt{Tortora:2013}). Gravitational lensing over galaxy scales provides a third,
independent, probe of the IMF in massive ETGs
\citep{Treu:10,SmithLucey2013}.

Subsequent studies focused on radial gradients of the IMF of ETGs
\citep{NMN:15a,NMN:15b,LB:16,LB:17,Ziel:17,vanDokkum:2017,Parikh:2018,Sarzi:2018}.
These studies revealed that the non-standard, bottom heavy IMF is only
found in the central regions of the most massive galaxies. At face
value, massive galaxies are therefore made up of two different types
of stellar populations, produced in substantially different modes of
star formation. This interpretation aligns with the paradigm of a
two-stage formation process \citep{Oser:10}, whereby the stellar
populations of massive galaxies can be split into a component formed
in-situ, mostly during the first, intense phases of formation, and an
ex-situ component contributed by mergers. Regarding the in-situ phase,
the massive ``red nuggets'' found at high redshift \citep[see,
  e.g.,][]{Daddi:2005, Trujillo:2007,vdK:08,Damjanov:2011}, suggest that in
these systems, a gas
mass $\gtrsim 10^{11}$M$_\odot$ must have been transformed into
stars within a R$\sim$1-2\,kpc region in, roughly, a dynamical time
($\sim$1\,Gyr), therefore leading to a sustained star formation rate
$\gtrsim$100\,M$_\odot$\,yr$^{-1}$. Such an extreme environment is
bound to produce highly supersonic turbulence in the gas, resulting in
fragmentation over small scales, therefore favouring a bottom-heavy
IMF (\citealt{PadNor:02,Hopkins:13,Chabrier:14}; but see 
also~\citealt{Bertelli:2016}).

Intriguingly, observations also find that in strongly star-forming
systems, the IMF is, rather, top-heavy, i.e. producing an excess of
high-mass stars with respect to the Milky Way standard
\citep{Gunaw:11}. This conundrum can be solved by invoking a
time-dependent IMF whereby a strongly star forming system starts with a
top-heavy IMF, followed by a later phase where a large amount of gas
is locked into low-mass stars~\citep{Vazdekis1996, Vazdekis1997, Dave:2008}. This scenario is admittedly contrived,
but it is shown to produce compatible chemical enrichment properties,
X-ray binary fractions, as well as the IMF signatures found in massive
ETGs \citep{Weidner2013,Ferreras2015}. The suggested claim of a
potential correlation between IMF and metallicity \citep[hereafter MN15b]{NMN:15b} 
gave rise to a number of modeling attempts to understand such
a trend \citep{Clauwens:16}, including detailed 
hydrodynamical simulations \citep{Gutcke:19} or the concept of the
Integrated Galactic IMF \citep{TJ:18}. Additional models have been
explored to shed light on alternative causes of IMF variations in
massive ETGs, such as cosmic rays \citep{Fontanot:18}, but the final say rests on
detailed, ab initio (magneto-)hydrodynamical simulations under the physical
conditions expected during the early formation of the massive cores of the
ETGs we see today.

In spite of the advances made over the past few years on the issue of line
strength/spectral fitting constraints on the IMF of ETGs, a number of
key issues remain controversial. The interpretation of the spectral
features is still open to debate, as, for instance, the spectral
indices targeted to discriminate between giants and low mass dwarfs in
evolved populations is prone to degeneracies, with respect to age, and
most notably, chemical composition. For instance, some claims were
made that the strong Na\,{\sc I}-dependent feature found at $\lambda\sim 0.82\mu$m\ 
could be partly, or even fully, described by an overabundant [Na/Fe]
\citep{Jeong:2013,McConnell:16}. Such a problem is overcome by the joint analysis of a 
battery of spectral features with different sensitivity to specific
chemical abundance ratios \citep[see, e.g.,][]{Spiniello:2012,LB:13}, and by a
detailed analysis of a number of Na-sensitive line strengths over a
wide spectral window \citep{LB:17}. Moreover, the finding of massive
ETGs with apparently standard IMFs, but selected as strong
gravitational lenses \citep{SmithLucey2013,smith:2015,SLC:2015,Leier:16} poses
additional questions as to why these galaxies behave 
differently to the general population of massive ETGs found in SDSS-based
samples involving thousands of spectra \citep{Newman:2017}.

This paper completes the study of a sample of massive ETGs observed
with the X-Shooter instrument at the European Southern
Observatory--Very Large Telescope~\citep{Vernet:2011}.  This project was aimed at
exploring in detail the radial gradients of the IMF in a carefully
defined set of ETGs with very high velocity dispersion -- where the
departure from a standard IMF is expected to be highest -- taking
advantage of the wide spectral coverage allowed by the X-Shooter
spectrograph, as well as the latest, state-of-the-art models of
stellar population synthesis, coupled to theoretical models of stellar
atmospheres to decipher the role of chemical composition in the
derivation of the IMF.  In \citet[hereafter LB16]{LB:16}, we analyzed
the radial IMF gradient of one of the targeted ETGs (XSG1),
based on TiO spectral features and
the Wing-Ford band, finding robust trends regarding the IMF, along
with further constraints on the shape of the mass function at the very
low mass end.  In \citet[hereafter LB17]{LB:17}, we extended the
analysis by including four Na-sensitive indices that cover a wide
spectral range (between $\lambda$$\sim$0.6 and 2.2\,$\mu$m), and targeted two ETGs,
confirming previous results of IMF variations regardless of the
[Na/Fe] chemistry. This paper presents a comprehensive analysis of the
full data set, comprising seven ETGs, with the full spectral range of
X-Shooter allowing us to target a comprehensive set of line strengths.

The paper is structured as follows.  In Sec.~\ref{sec:specdata}, we
describe our sample of massive ETGs (hereafter XSG sample), and the
new X-Shooter data.  
Sec.~\ref{sec:spmodels}
describes the stellar population models used to analyze the spectra.
The approach to perform the stellar population study is detailed in Sec.~\ref{sec:analysis}, while Sec.~\ref{sec:results} presents the results. A discussion follows in Sec.~\ref{sec:Discussion}.

%
%
%


\section{Sample and New Data}
\label{sec:specdata}

Our sample consists of seven massive ETGs. Six galaxies have been selected from the pool of most massive 
ETGs at the lowest redshift limit ($z \sim 0.05$) of the SPIDER survey~\citep[hereafter SpiderI]{SpiderI}, 
while one extra ETG (named XSG10, see below) has been selected from SDSS-DR7 applying the same criteria as SPIDER ETGS, but at slightly lower redshift ($z \sim 0.048$) than the SPIDER sample ($0.05 \! \le \! z \! \le  \! 0.095$). 
In F13 and LB13, we have used the highest-quality SDSS spectra of SPIDER ETGs to construct a subsample of 18 stacked spectra~\footnote{Notice that the 18 stacked spectra analyzed in F13 and LB13 are made public available at the following link: http://www.iac.es/proyecto/miles/pages/other-predictionsdata.php. }, within narrow bins of central velocity dispersion, 
$\sigma$, each bin with a width of $\rm 10$~\kms , except for the last two, with $260 \! \le \! \sigma \! \le \! 280$, and $280 \! \le \! \sigma \! \le \! 320$~\kms, respectively.
All targets (including XSG10) analyzed in the present work have $\sigma$  in the range of the highest velocity dispersion bin (from $280$ to $320\,$\kms) defined in LB13, { and have been selected based on line-strengths of IMF-sensitive absorption features  and abundance ratio estimates from the SDSS spectra. In particular, in order to probe the range of values in the parent $\sigma$ bin, ETGs in our sample have SDSS estimates of \mgfe\ in the range from $\sim 0.3$ to $\sim 0.5$~dex (with a typical uncertainty of 0.1~dex; see LB13), and SDSS measurments of \tioii\ ( \nai ) in the range from $\sim 0.08$ to $\sim 0.095$~mag ($4.5$ to $5.5$~\AA). 
Therefore, our sample should be representative of the high-mass end population of ETGs.} Throughout the present work, we refer to our targets 
as XSG1, XSG2, XSG6~\footnote{Notice that other three targets, named XSG3, XSG4, and XSG5, were also included in our VLT X-Shooter observing proposals, but not observed. Therefore, they are not included in out sample.  }, XSG7, XSG8, XSG9, and XSG10, respectively. Basic galaxy properties, including the SDSS identification of each object, are provided in Tab.~\ref{tab:galaxies}. In App.~\ref{app:envi}, we describe the environment where the XSGs reside, based on the SDSS-DR7 group catalogue of~\citet{Wang:2014}. Most of the XSGs, i.e. XSG1, XSG6, XSG8, XSG9, and XSG10 are centrals of a galaxy group, while XSG2 and XSG7 are satellites~\footnote{However, as discussed in the Appendix, the classification of XSG7 is uncertain, this galaxy having similar mass of, and being very close to, the brightest group galaxy.}. Also, in App.~\ref{sec:surfphotometry}, we present a surface photometry analysis of the XSGs, based on the SDSS photometry. 
Remarkably, our analysis shows that the estimate of the effective radius, $\rm R_e$, for most XSGs depends significantly on the method used to estimate such a quantity. Since we cannot assign a unique scale-length to XSGs' light profiles, when studying stellar population properties as a function of radius (see below), we consider galactocentric distances in units of kpc, without rescaling them by (a given estimate of) the effective radius. We also discuss the effect of rescaling the IMF profiles with $\rm R_e$  in App.~\ref{app:radii}, and we come back on this point in Sec.~\ref{sec:results}.

\begin{table*}
\centering
\small
 \caption{Main galaxy properties. Column~1 gives the label used throughout the present work for each galaxy.
 Column~2  is the galaxy SDSS ID, while columns 3 and 4
 are the galaxy RA and DEC coordinates. Column~5 is the galaxy total magnitude in r band, obtained by averaging total magnitudes of best-fitting Sersic and B+D  models (see Sec.~\ref{sec:spmodels} for details). Columns 6 and~7 report galaxy redshifts, as derived 
 from our X-Shooter spectroscopy, and retrieved from the SDSS database, respectively. Columns~8 and~9 report SDSS and X-Shooter central velocity dispersions of each galaxy, respectively. }
  \begin{tabular}{c|c|c|c|c|c|c|c|c}
   \hline
XSG\# &  SDSS ID & RA & DEC & $\rm M_r$ & z & z & $\rm \sigma_0$ & $\rm \sigma_0$ \\
      &   &  &   & & SDSS & X-Shooter & SDSS & X-Shooter \\
     &   & (deg) & (deg)  & (mag) & &  & (\kms) & (\kms) \\
   (1)  &  (2) & (3) & (4) &  (5) & (6) & (7) & (8) & (9) \\
  \hline
1  & J142940.63+002159   & 217.41929 &  0.366398 & $-22.7$ & $0.055787 $ &  $0.055757 $& $301 \pm 9  $ & $333\pm 3$ \\
& & & & & $\pm 0.0002$ &$\pm 0.000006$ &  & \\
2  & J002819.3-001446.7  &   7.08043 & -0.246338 & $-22.3$ & $0.059951 $ &  $0.05992 $ & $292 \pm 11 $ & $302 \pm 9$ \\
& & & & & $\pm 0.0002$ &$\pm 0.000007$ &  & \\
6  & J144120.36+104749.8 & 220.33484 &  10.79719 & $-23.3$ & $0.051279 $ &  $0.051249 $& $286 \pm 10 $ & $305 \pm 13$ \\
& & & & & $\pm 0.0002$ &$\pm 0.000008$ &  & \\
7  & J151451.68+101530.4 & 228.71533 &  10.25845 & $-22.7$ & $0.054857 $ &  $0.054843 $& $288 \pm 12 $ & $319 \pm 15$ \\
& & & & & $\pm 0.0002$ &$\pm 0.000009$ &  & \\
8  & J015418.07-094248.4 & 28.575312 &  -9.71347 & $-23.8$ & $0.052451 $ &  $0.052353 $& $293 \pm 11 $ & $332 \pm 8$ \\
& & & & & $\pm 0.0002$ &$\pm 0.000007$ &  & \\
9  & J005551.88-095908.3 & 13.966206 &  -9.98565 & $-23.7$ & $0.054750 $ &  $0.05476  $& $296 \pm 12 $ & $349 \pm 6$ \\
& & & & & $\pm 0.0002$ &$\pm 0.000008$ &  & \\
10 & J075354.98+130916.5 &118.479088 &  13.15459 & $-22.5$ & $0.047671 $ &  $0.04763  $& $305 \pm 10 $ & $338 \pm 4$ \\
& & & & & $\pm 0.0002$ &$\pm 0.000006$ &  & \\
\hline
  \end{tabular}
\label{tab:galaxies}
\end{table*}

\subsection{Observations and data reduction}
\label{subsubsec:data}

For all seven targets, we have obtained new, deep long-slit spectroscopy
with the X-Shooter spectrograph at the ESO-VLT, on Cerro Paranal
(Proposal IDs: 092.B-0378, 094.B-0747; 097.B-0229; PI: FLB).  X-Shooter is a
second-generation ESO-VLT instrument -- a slit echelle spectrograph
that covers a wide spectral range (3000--25000\,\AA), at relatively
high resolution~\citep{Vernet:2011}. The spectral range is covered by splitting the incoming bin into three independent arms, ultraviolet-blue (UVB: 3000--5900\,\AA);
visible (VIS: 5300--10200\,\AA); and near-infrared (NIR: 9800--25000\,\AA).
Details on the data for XSG1 and XSG2 have been provided in LB16 and LB17. We give 
here only a short summary.
The X-Shooter slit is 11'' long, with a spatial scale of 0.16\,$\rm
arcsec/pixel$ in the UVB and VIS, and 0.21\,$\rm arcsec/pixel$ in the
NIR, arms. For all observations, we adopted an instrument setup with 0.9''-, 0.9''-, and 1.0''- wide slits, resulting into a resolution power of 
$R \sim\!4400$, $\sim\!7500$, and $\sim\!5500$, in the UVB, VIS, and NIR arms,
respectively.
We have observed XSG1 (XSG2) through five (ten) observing
blocks (OBs), each including two exposures on target, interspersed by
two (one) sky exposures, with the same integration time as for the
science target.  This setup gives a total on-target exposure
time of $\sim$1.7, 1.9, and $2.1$\,hr, in the UVB, VIS, and
NIR arms, respectively (see LB16 and LB17 for details).
XSG6, XSG7, XSG8, XSG9, and XSG10 were observed with the same setup as for XSG1.
In order to minimize slit losses due to problems with the X-Shooter atmospheric dispersion corrector, observations were taken at parallactic angle, resulting into data taken at two/three position angles (depending on the galaxy) for each target.
Observations were carried out in service mode, with a median seeing (as measured at the telescope) of $\sim$0.8--0.9$''$ (FWHM), depending on target. Due to bad weather conditions, one (two) exposure(s) 
for XSG6 (XSG7) were not usable for our purposes, 
resulting into a slightly lower exposure time for these galaxies. 
For each arm, the data were pre-reduced using
version 2.4.0 of the data-reduction pipeline~\citep{Mod:2010},
performing the subsequent reduction steps (i.e. flux calibration, sky
subtraction, and telluric correction) with dedicated FORTRAN software
developed by the authors. We refer the reader to LB16 and LB17 for a detailed
description of each reduction step (see also~\citealt{SCH:2014}).
As shown in App.~\ref{app:SDSS}, in the galaxy central regions, our new X-Shooter data are fully consistent with the existing (optical) SDSS spectroscopy. 

\subsection{Radial binning}
\label{subsubsec:radialbins}
For each galaxy we extracted 1D spectra at different galactocentric distances,
by summing up all  available exposures, and folding up data
from opposite sides of the X-Shooter slit around the galaxy photometric
centre (see LB16 for details). For the latter step, each row of the  two-dimensional spectrum was 
first corrected to restframe, using the rotation velocity profile of each galaxy, as derived with 
the software {\sc pPXF}~\citep[see below]{Cap:2004}.
The kinematics of the XSGs will be presented in a forthcoming paper. For the present work, we just notice that
all galaxies  have little rotation velocity, less than $\sim 50$ ($\sim 100$)\kms\ for XSG6, XSG7, XSG8, XSG9, and XSG10 (XSG1 and XSG2; see figure~2 of LB16 for the kinematic profile of XSG1).
In order to minimize seeing effects, the innermost spectra of all galaxies were extracted within an aperture of 
width 1.3$^{\prime\prime}$ (i.e. $\pm 0.675^{\prime\prime}$) around the photometric centre,
corresponding to a factor of 1.5 times the mean seeing FWHM of our observations. The bin size was 
then increased adaptively outwards, in order to ensure an high  median S/N ($>$90) per \AA\ in 
the optical spectral range (from 4800 to 5600\,\AA).
This procedure gives six radially binned spectra for XSG1 and XSG8; five binned spectra 
for XSG6, XSG7, XSG9, XSG10; and four radial bins for XSG2.
For XSG1, the binned spectra have been shown in figure~4 of LB16, with spectra for all the other galaxies having 
similar quality. In particular, the median S/N measured in the central bins is very high, with values ranging from 170 (per \AA ), for XSG9, to 270 for XSG10.

For all radially binned spectra,  the velocity dispersion $\sigma$ was measured 
as detailed in LB16, by running {\sc pPXF} on different spectral regions of the 
UVB and VIS X-Shooter arms ($\lambda\lambda =4000-9000$\,\AA ), and combining the corresponding probability distribution functions into final estimates. 
For each galaxy, the velocity dispersion of the central X-Shooter spectrum, $\sigma_0$, is reported in Tab.~\ref{tab:galaxies}, together with the 
SDSS measurements of $\sigma$.
The quoted uncertainties on $\sigma_0$ combine, in quadrature, the formal measurement errors with the rms of $\sigma_0$ estimates among different  spectral regions (see above).
All the XSG's have $\sigma_0$ larger than $300$~\kms, consistent with the selection of these objects to be very massive  
systems. The velocity dispersion profiles will be presented in a forthcoming paper. We just point out here that all galaxies have a shallow $\sigma$ gradient with galactocentric distance, with a drop of $\sim 40$--$70$~\kms\ (depending on the galaxy), 
over the radial range probed by X-Shooter. Because of that, the SDSS values of $\sigma_0$ are  smaller (by $\sim 10 \%$, on average) than those estimated with X-Shooter (see the Table), as they refer to the SDSS fiber aperture of radius $1.5 ''$, a factor
more than two larger than the size of the X-Shooter innermost radial bins (see above).

\section{Stellar population models}
\label{sec:spmodels}

\subsection{Na--EMILES SSPs}
\label{subsec:MILESmodels}
Our analysis relies on the Na--EMILES stellar population models, a dedicated 
version  of the EMILES models covering a range of \nafe\ abundance 
ratios (see LB17). EMILES models cover the
spectral range from $0.35$ to $5 \mu$m, at ``moderately'' high
resolution (see below). Such wide wavelength range is achieved by joining
different simple stellar population (SSP) model predictions based on 
empirical stellar libraries, namely MILES in the optical range \citep{MILESI}, 
from $\lambda \! \sim \! 3540$\AA\ to $\lambda \! \sim \! 7410$\AA , Indo-US~\citep{Valdes04} and 
CaT~\citep{CATI} out to $\lambda \! \sim \! 8950$~\AA\ (\citealt{Vazdekis:12}), and 
the IRTF stellar library~\citep{IRTFI,IRTFII}, extending the $\lambda$ range out 
to 5\,$\mu$m (see \citealt{RV:16} for details about the joining procedure). 
The spectral resolution is kept constant with wavelength (at FWHM=2.5\AA) for all libraries, except for IRTF, having a constant $\sigma$=60~\kms\ (see figure~8 of~\citealt{Vazdekis:2016}).
The models are computed for two sets of scaled-solar theoretical isochrones, namely
the ones of \citet{Pietrinferni04} (BaSTI) and \citet{Padova00}
(Padova00).  The BaSTI isochrones are supplemented with the stellar
models of \citet{Cassisi00}, which allow the very low-mass
(VLM) regime to be covered down to $0.1 \, M_\odot$. The temperatures of these low-mass
stars are cooler than those of Padova00 \citep{Vazdekis:12}. 
Both sets of isochrones include the thermally pulsing AGB regime using simple synthetic prescriptions,
providing a significantly smaller contribution for this evolutionary
phase at intermediate-aged stellar populations in comparison to the
models of \citet{Marigo08} and \citet{Maraston05}, and more similar to that of~\citet{BC03} models. Finally, there are
differences between the BaSTI and Padova00 isochrones, described in
detail in \citet{Cassisietal04}, \citet{Pietrinferni04}, and~\citet[hereafter V15]{Vazdekis:15}.
To cover a range in \nafe , we apply theoretical differential corrections for \nafe\ overabundance,
specifically computed for each individual stellar spectrum in the
empirical libraries. The Na--EMILES SSPs are then constructed, as detailed in LB17, based 
on scaled-solar isochrones. Notice that this approach differs from that used in V15, where 
we have computed 
\alfa-- (rather than Na-- ) enhanced models based on corrections applied directly to the 
model SSPs, rather than to individual stars. Also, the \alfa--enhanced
models only cover the optical (MILES) spectral range, while Na--enhanced models are computed over the optical plus NIR spectral range (see above). We point out that the Na--EMILES models used in the present 
analysis differ from those presented in our previous work, in that (i) we have constructed both 
BaSTI and Padova00 SSPs (rather than only Padova00 models, as in LB17), (ii) the models are computed for a wider range of IMF slopes (see below); and (iii) stellar spectra have been corrected to \mgfe$=0$, based on \mgfe\ abundances from~\citet{Milone:2011}, using theoretical differential corrections computed for each individual stellar spectrum in the empirical libraries~\footnote{However, since  the abundance pattern of stars in the empirical libraries follows that of the Galaxy (i.e. stars are alpha-enhanced at sub-solar metallicity, while they have \mgfe$\sim 0$ in the solar and super-solar metallicity regime),  the corrections have negligible impacts on models relevant for the present work, where we study massive ETGs, whose spectra typically have super-solar metallicities.}.

Na--EMILES models are computed for different IMF shapes, as described in
\citet{CATIV} and V15, and in particular for two power-law distributions,
as defined in \citet{Vazdekis1996}, i.e unimodal (single power-law) and
bimodal. The lower and upper mass-cutoffs are set to $0.1$ and
$100$\,M$_\odot$, respectively. The unimodal and bimodal IMFs are
defined by their logarithmic slope, $\Gamma$ and $\Gamma_b$,
respectively. For reference, the \citet{Salpeter55} IMF is obtained when
adopting a unimodal IMF with $\Gamma=1.35$, whereas the~\citet{Kroupa01}
Universal IMF is closely approximated by a bimodal IMF with
$\Gamma_b=1.3$. The difference between unimodal and bimodal IMFs is that the bimodal distribution
is smoothly tapered towards low masses ($\sim 0.5 \, M_\odot$); hence, varying  the
high-mass end slope $\Gamma_b$ changes the dwarf-to-giant ratio in the IMF through its overall
normalization. While this approach is different with respect to a
change of the low-mass slope (e.g.~\citealt{CvD12b}), the bimodal parameterisation is
suitable for our purposes, as most IMF-sensitive features depend on the dwarf-to-giant ratio in the IMF (e.g.~LB13,
LB16). 
Since a low-mass tapered IMF provides mass-to-light ratios consistent with
dynamical constraints~\citep{Lyubenova} and is able to describe both optical and NIR 
IMF-sensitive features (see LB16, where the fits with a single power-law distribution were largely disfavoured), in the present work we consider only models with a low-mass tapered distribution. 

\subsection{Spectral indices}
\label{subsec:indices}

\begin{figure*}
 \begin{center}
\leavevmode
    \includegraphics[width=15cm]{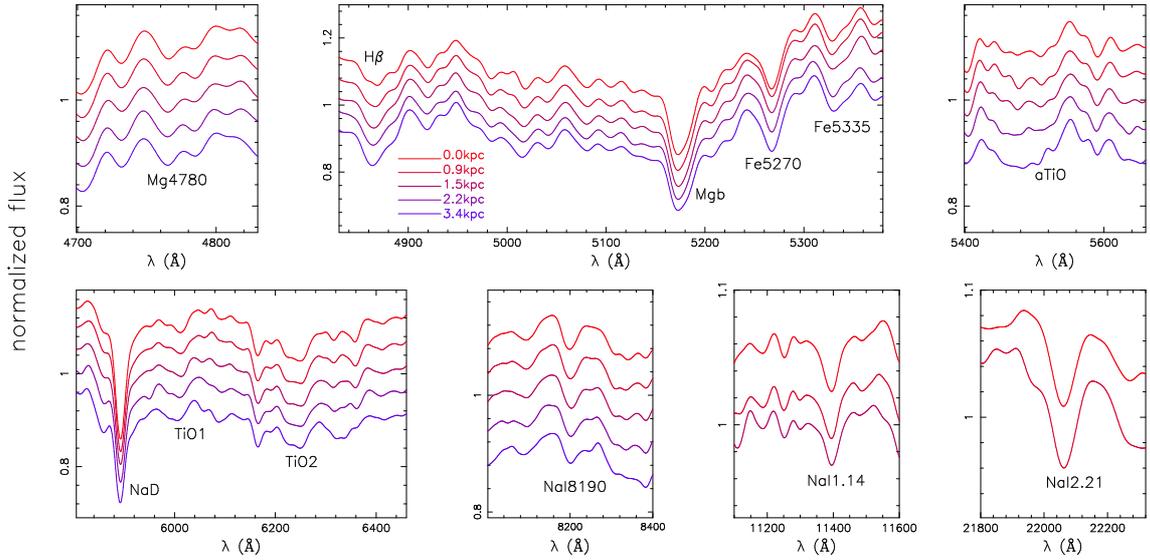}
 \end{center}
    \caption{ X-SHOOTER spectra of XSG6 (one of the new galaxies analyzed in the present work). Different panels show different spectral regions, with the main absorption features included in our analysis (see labels in each plot, and Sec.~\ref{subsec:fitting}). 
    The data are radially binned in five galacto-centric distances, out to $\rm
  R\sim 3.5$~kpc, from the galaxy center. Different colours, from red through blue,
  correspond to different radial positions, as labelled in the top-middle panel.  All spectra have been smoothed, for
  displaying purposes, to a common velocity dispersion of 400\,\kms . Notice that \naiii\ and \naiv\ are only included in the analysis for  the innermost radial bins of each galaxy, as in the outermost spectra these indices have large uncertainties and/or are affected by sky subtraction residuals (see Sec.~\ref{subsec:indices}).
    }
    \label{fig:XSG6spec}
\end{figure*}

The wide spectral range provided by X-Shooter allows us to analyze both optical and NIR spectral features simultaneously. Following the same approach as in our previous works (e.g. LB13, LB16, LB17), we constrain stellar population properties by comparing observed and model line-strengths for a selected set of optical and NIR spectral indices. For the present work, we consider the age-sensitive Balmer lines, \hbo\ and \hgf, the total metallicity indicator \mgfep , the IMF-sensitive features \tioi, \tioiio, \atio, and \mgf, as well as the four Na indices, \nai, \naii, \naiii, and \naiv~\footnote{
 Notice that, for each spectrum,  only indices not affected by sky residuals, and/or large uncertainties ($> 100 \%$) were included in the analysis. For this reason, \naiv\ is only included in the analysis for  the innermost radial bins of each galaxy (see App.~\ref{app:bestfitindices}).
}, which are sensitive to both IMF and Na abundance ratio. 
The index definition for \tioi, \hgf, and \nai\ is the same as in \citet{Trager98}, while \hbo\ is the optimized \hb\ index defined by \citet{CV09}. The \tioiio\ is defined as in LB13, being a modified version of \tioii\ from \citet{Trager98}. 
The total-metallicity indicator \mgfep\ is a combined Mgb and Fe index, defined by~\citet{TMB:03} to be insensitive to \mgfe\ abundance ratios (see also V15). Finally the \mgf\ is from~\citet{Serven:2005}, \atio\ from~\citet{Spiniello:2014}, while \naii, \naiii, and \naiv\ are defined as in \citet[hereafter CvD12a]{CvD12a}, with some modifications (in the air system) as described in LB17.
In a forthcoming paper, we will also explore features from other chemical species (e.g. Ca, K, CO) available over the wide X-Shooter spectral range, thanks to the ongoing development of dedicated stellar population models (as the Na--EMILES).
{ Fig.~\ref{fig:XSG6spec} shows, as an example, the radially binned spectra for one galaxy in our sample, with the inset panels zooming into the spectral regions around the selected IMF-sensitive spectral features.}

For each IMF slope (see Sec.~\ref{subsec:MILESmodels}), we compute predictions for single SSP models (hereafter 1SSP) as well as models with a fraction of young stars added on top of an old SSP (hereafter 2SSP).   In the 1SSP case, we compute indices for SSPs with ages between about $4$ and $14$~Gyr, and metallicities between about $-0.4$ and $+0.2$~dex, as younger/lower-metallicity SSPs are not relevant for our sample of massive ETGs. More specifically, in the case of Padova00 models, we use SSPs with ages of \{3.5481, 5.0119, 6.3096, 7.9433, 8.9125, 10.0000, 11.2202, 12.5893, 14.1254\}~Gyr, and  metallicites \{-0.4, 0, +0.22 \}~dex. For BaSTI models, the adopted SSPs 
have ages in the range from 4 to 14~Gyr,
at  a step of 1~Gyr, and metallicites \{-0.35, -0.25, +0.06, +0.15, +0.26\}~dex. Notice that BaSTI models are computed for more metallicities than the Padova00 set (see V15). In particular, BaSTI SSPs are also computed for \zh=0.4~dex, but these models are unsafe according to the quality parameter defined in V15, and therefore are not used in the present work.  
The computed line-strengths are linearly extrapolated~\footnote{The extrapolation is carried out by performing a linear fit of each index versus the total metallicity indicator \mgfep . The fit is done by considering only model predictions with \zh$\ge 0$, for both Padova00 and BaSTI models. 
} out to \zh=0.7 for Padova00, and out to 0.75~dex~\footnote{In practice, if we consider Padova00 models, only a few spectra (i.e. the three innermost bins for XSG8, and the outermost bin for XSG6) require a significant amount of extrapolation, while for all the other cases, metallicity is within the range of the models (\zh$\le +0.22$), or requires a small amount of extrapolation (by less than $\sim $0.1~dex). Therefore, none of our conclusions is affected by the extrapolation procedure. } 
for BaSTI models, and then interpolated over finer grids with 111 (42) steps in age (metallicity).
All index grids are computed for different values of \nafe=\{0, 0.3, 0.6, 0.9\}, for which Na--EMILES models have been computed. Models with \nafe=1.2 are also computed, as in LB16, but are not used for the present analysis, as we verified {\it a posteriori} that none of our galaxy requires such extreme \nafe\ abundance ratios (see Sec.~\ref{subsec:IMFdrivers}). Index grids are finally interpolated with a finer step of 0.01~dex in \nafe. 

In the 2SSP case, we compute indices for linear combinations of two SSPs, with  the ``old'' SSP having the same values of age, metallicity, IMF slope, and \nafe\, as in the 1SSP case, and a younger SSP, whose age is varied between 1 and 4~Gyr, for three metallicities of \{-0.4, 0, +0.2\} (\{-0.35, +0.06, +0.26\}) for Padova00 (BaSTI) models. The light fraction of the young to old component is varied between 1 to 15~\%~\footnote{We verified, a posteriori, that none of our fits requires a 
fraction of young-to-old stars larger  than 15~\%.}.

Notice that in order to maximize the information provided by the data, we do not smooth all spectra and models at the same sigma, but instead,  for each spectrum (i.e. each galaxy/radial bin),  we compare observed line-strengths to model predictions computed at the same sigma as the given spectrum (see Sec.~\ref{subsec:IMFdrivers}).

\section{Analysis}
\label{sec:analysis}

\subsection{Fitting procedure}
\label{subsec:fitting}

To constrain the IMF, we adopt a similar approach to that in LB16, updated to fit Na absorption lines as in LB17. Overall, the method  follows the approach described in our previous works (e.g.~F13, LB13,~\citealt{LB:15, NMN:15a}, MN15b,~\citealt{NMN:15c}).  For each spectrum, we minimize the following expression,
\begin{eqnarray}
\rm \chi^2(Age, {\rm [Z/H]}, \Gamma_b, [Na/Fe]) = 
  \rm \sum_i \left[ \frac{E_{obs,i}^{ss}- E_{mod,i} }{\sigma_{E_{obs,i}^{ss}}} \right]^2 + \nonumber \\
  \rm \sum_{j} (  Na_{obs,j} - Na_{mod,j}  
         -   \alpha_{Na_j} \cdot Na_{mod,j} \cdot [\alpha/Fe] )^2 \cdot {\sigma_{j}}^{-2} 
\label{eq:method}
\end{eqnarray} 
where the index $j$ runs over Na features, while the 
index $i$ runs over the remaining selected  spectral features (see Sec.~\ref{subsec:indices} and Sec.~\ref{subsec:fits}); 
$\rm E_{mod,i}$ and $\rm Na_{mod,j}$ are line-strength predictions for a given (either 1SSP or 2SSP) model, $\rm E_{obs,i}^{ss}$ are observed line-strengths -- $\rm E_{obs,i}$ --
corrected to solar-scale as detailed in LB16, while $\rm Na_{obs,j}$ are observed line-strengths for Na features. The \afe\ is the \alfa-element abundance ratio for the given spectrum, estimated through the solar-scale proxy \afep\ (see LB13 for details), while \alj\ is the {\it normalized response} ($\rm \delta Na_J / Na_j$) of the j-th Na-sensitive index to \afe. The \alj\ have been constrained, { empirically}, in LB17 (see figure~5 in 
that paper), being consistent, in the optical, with theoretical predictions from \alfa--MILES models. The terms $\rm \sigma_{E_{obs,i}^{ss}}$ and $\rm \sigma_{j}$ are the uncertainties on solar-scale corrected line-strengths and Na line-strengths, respectively. The minimization of Eq.~\ref{eq:method} is performed over a discrete index grid of Na--EMILES model predictions, with varying Age and \zh\ (for 1SSP models), \gammab\, and \nafe\ abundance ratios. 
For 2SSP models, the grid includes three extra parameters, given by the Age and \zh\ of the young component, as well as the relative light fraction of the young vs. old components (see Sec.~\ref{sec:spmodels}).
In the model grid, the effect of varying \nafe\ is considered only for Na features, as for the other features, the (overall) effect of abundance ratios is already taken into account by the empirical corrections. Uncertainties on best-fitting parameters, \{Age, ${\rm [Z/H]}$, \gammab, \nafe \}, are obtained from $N=1000$ bootstrap iterations, where the fitting is repeated after
shifting observed line strengths (as well as the \alj's) according to their uncertainties.

Notice that, besides the effect of age, metallicity, and IMF slope,  Na features  are mostly sensitive only to \nafe\ and \afe\ abundance ratios (see CvD12a, LB17; but see also~\citealt{Benny:2017} for a possible contribution of \cfe\ to \naiv ), whose effects are taken into account by Na--EMILES 
models and the \alj\ coefficients (from LB17), respectively. Moreover, as shown in LB17, the effects of \nafe\ and IMF on Na features are actually coupled, with a bottom-heavy IMF boosting up the \nafe\ response of the (NIR) $\rm Na_j$'s. This coupling is also taken into account by our models, where Na-enhanced SSPs are computed for all different IMF's.
For these reasons, in Eq.~\ref{eq:method}, Na features are treated in a different manner with respect to the other features, where the (cumulative) effect of abundance ratios (from different chemical species) is taken into account by our empirical correction procedure. Notice, however, 
that as discussed in LB16, the combined effects of abundance ratios is very small for most of the features (e.g. the TiO's) as the effect of different elements (e.g. \afe\ and \cfe ) compensate each other. Therefore, our results are not driven by the empirical correction approach. 

\subsection{Fitting schemes}
\label{subsec:fits}

In order to check the robustness of our results, we consider different fitting cases, by (i) changing  the set of fitted indices, (ii) exploring both 1SSP and 2SSP models, (iii) using models based on different isochrones, i.e. both BaSTI and Padova00 Na--EMILES. The different cases are summarized in Tab.~\ref{tab:methods}. Case A fits 1SSP Padova00 models, including all indices (but the Balmer lines~\footnote{
Balmer lines (in particular \hb ) are also sensitive to IMF, individual abundance ratios (e.g. \cfe ), and small fractions of young stars (see LB13, LB16, and references therein). Hence, any age determination based on \hb\ might bias the (simultaneous) inference of IMF slope. For this reason, we have not included \hb\ in method A (although we verified that including it does not affect significantly our results), but we have fully exploited the effect of fitting Balmer lines through method D, as detailed in the text.}), while cases B and C~\footnote{Notice that in case C, Eq.~\ref{eq:method} does not include the term with Na line-strengths.} are the same as A but removing different sets of indices, i.e. the TiO's and all indices but TiO1 and TiO2, respectively.  Case D is the same as A but for two SSPs, including also \hb\ and \hgamma\ Balmer lines. Notice that in case A we do not include \hb\ in the fit, as this index is also sensitive to IMF (see LB13), and we want to minimize any correlated variation of age and IMF slope. The effect of including the Balmer lines is explored with method D (providing very consistent results with method A, as shown in the following). Finally, case E is the same as A but changing the isochrones (i.e. BaSTI vs. Padova00). The rationale behind methods B and C is that TiO1 and TiO2 (i) are wide spectroscopic features, potentially affected by flux calibration issues (see LB16); (ii) have a prominent sensitivity to abundance ratios, such as [Ti/Fe], [O/Fe], and \cfe , affecting to less extent the other indices; and (iii) are more sensitive to temperature, than gravity (see~\citealt{Spiniello:2014}). Therefore, it is worth testing the impact of including/removing these indices from the analysis. { In particular, we notice that, since TiO1 and TiO2 are sensitive to \tife\ (with TiO2 being less sensitive to \tife\ than IMF, compared to TiO1) , while \atio\ and \mgf\ are not (the former  being mostly sensitive to $\rm [Fe/H]$, the latter to \mgfe\ and \nafe ; see LB16), the fitting scheme B is insensitive to non-solar \tife\ abundance ratios.}
To be consistent with our previous works (e.g. LB13 and LB16), we refer to case A as our ``reference'' fitting method, although none of the results are found to depend significantly on the fitting scheme.

\begin{table*}
\centering
\small
 \caption{Summary of different methods used to infer the IMF slope from observed line-strengths. Column 1 provides the label used to refer to each method. Column 2 reports the number of SSPs considered in the fitting procedure, while column 3 reports the models' isochrones,  with labels iP and iT referring to Padova00 and BaSTI isochrones, respectively (see Sec.\ref{sec:spmodels}). Column 4 gives the list of indices used for each method.}
  \begin{tabular}{c|c|c|c}
   \hline
 Method &  number of SSPs & isochrones & Spectral indices \\
   (1)  &     (2)      &  (3) & (4)     \\
  \hline
    A &  1 & iP & \mgfep , \tioi, \tioiio, \atio, \mgf, \nad, \naii, \naiii, \naiv \\
    B &  1 & iP & same as method A but w/o \tioi\ and \tioiio \\
    C &  1 & iP & \mgfep , \tioi, \tioiio\ (i.e. same as A but w/o \atio, \mgf, and Na indices) \\
    D &  2 & iP & same as method A plus \hbo\ and \hgf \\
    E &  1 & iT & same as method A \\
   \hline
  \end{tabular}
\label{tab:methods}
\end{table*}

\section{Results}
\label{sec:results}

\subsection{IMF radial gradients}
\label{subsec:IMFgradients}

\begin{figure*}
 \begin{center}
\leavevmode
    \includegraphics[width=11cm]{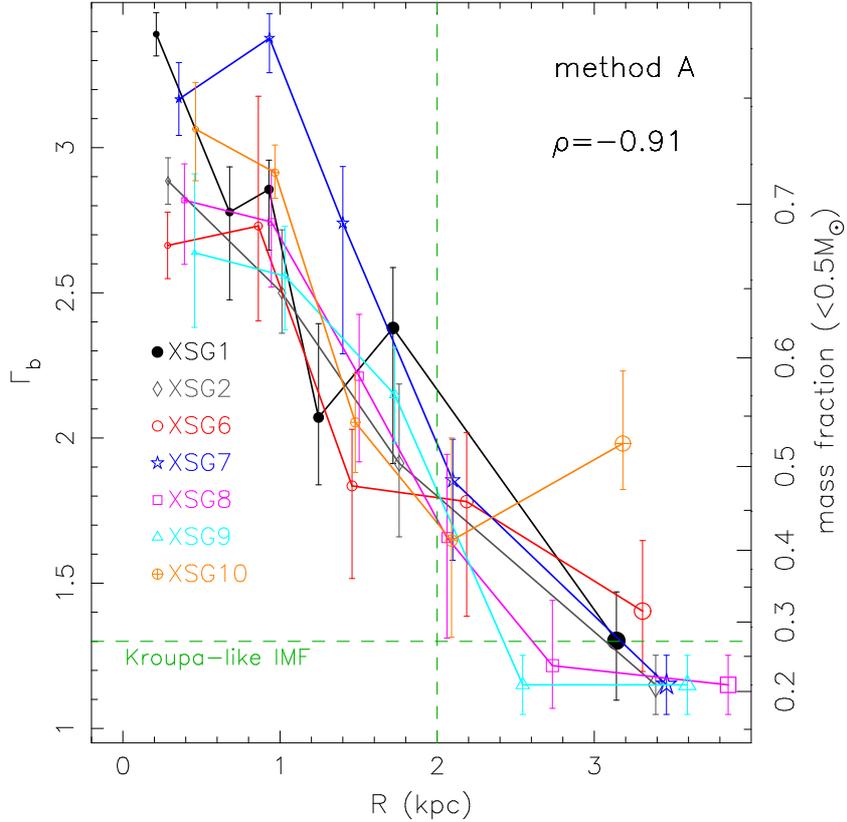}
 \end{center}
    \caption{IMF radial gradients in our sample of ETGs, for method A (i.e.  including all IMF-sensitive spectral indices in the fitting, and the same stellar population models as in LB16 and LB17; see Tab.~\ref{tab:methods}). 
    The y-axes report the IMF slope \gammab\ (left) and the mass fraction of low-mass ($\rm < M_\odot$) stars in the IMF, as defined in LB13 (see equation~4 of that paper).
    Different galaxies are plotted with different colors and line types, as labeled.
    Symbol sizes increase with galactocentric distance.
    Error bars denote 1$\sigma$ uncertainties.
    An IMF radial gradient is found for all galaxies, with a Kroupa-like IMF (\gammab$\sim 1.3$; see horizontal dashed line) at large radii ($\gtrsim 2$~kpc),
    and a bottom-heavy distribution in the center (leftwards the vertical dashed line). In the upper-right, we report the Pearson's correlation coefficient, $\rho$,  of \gammab\ versus R for the whole data-set (i.e. all galaxies and radial bins).
    }
    \label{fig:IMF_ref}
\end{figure*}

\begin{figure*}
 \begin{center}
\leavevmode
    \includegraphics[width=11cm]{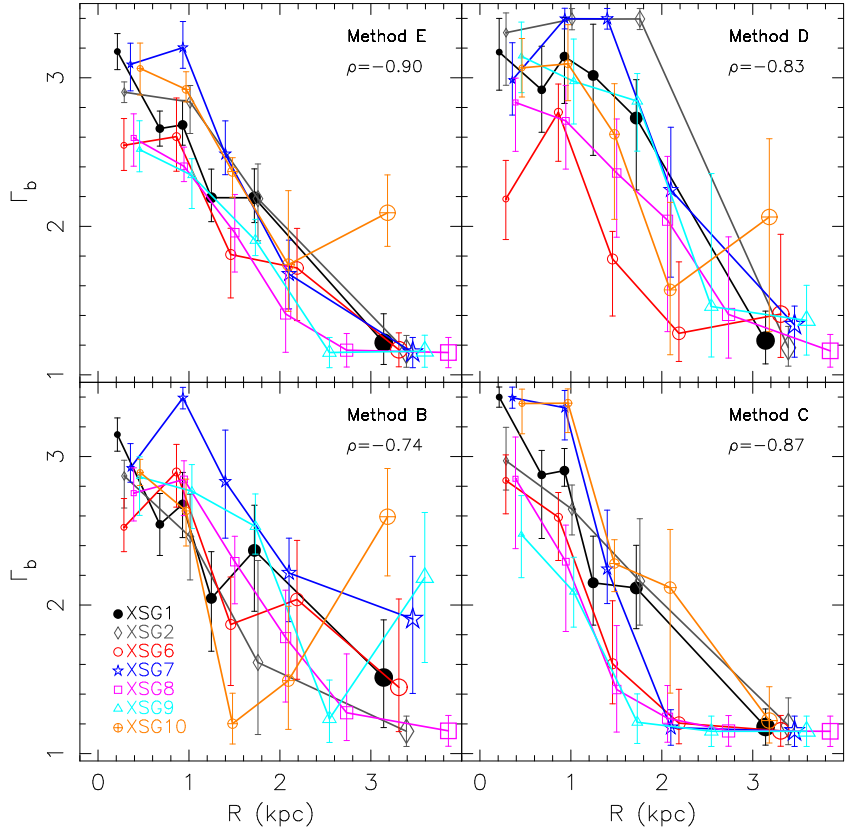}
 \end{center}
    \caption{IMF radial gradients for different methods, i.e. different sets of spectral indices and different models (see Tab.~\ref{tab:methods}). For each panel, different galaxies are plotted with different colors and line types (as labeled in the lower--left panel), with  
    symbol sizes increasing with galactocentric distance as in Fig.~\ref{fig:IMF_ref}.
    Upper--left: same as our ``reference'' method A (i.e. fitting 1SSP models, with all IMF-sensitive indices as well as the total metallicity indicator \mgfep ; see Fig.~\ref{fig:IMF_ref}), but based on models with BaSTI rather than Padova00 isochrones.  Upper--right: same as method A, but for 2SSP models, with Balmer lines included in the fitting procedure. Lower--left: same as method A, but excluding \tioi\ and \tioiio\ indices. Lower--right:  same as method A, but removing aTiO, Mg4780, and Na indices from the fitting (i.e. using only \mgfep , \tioi\ and \tioiio ). In the upper-right of each panel, we report the Pearson's correlation coefficient, $\rho$,  of \gammab\ versus R for the whole data-set (i.e. all galaxies and radial bins). Notice that, for all methods, an IMF radial gradient is detected. 
    }
    \label{fig:IMF_methods}
\end{figure*}

Fig.~\ref{fig:IMF_ref} plots the main result of the present work, i.e. the trend of IMF slope, \gammab, as a function of galactocentric distance, $\rm R$, in our sample of massive ETGs.  Results are plotted for  our reference fitting method (case A; see Tab.~\ref{tab:methods}). The values of \gammab$\sim 1.3$ correspond to a Kroupa-like IMF, and are marked with an horizontal dashed line in the Figure. All the XSGs show a negative IMF radial gradient, with a  bottom-heavy distribution in the galaxy central regions ($R \lesssim 2$~kpc; see vertical dashed line in the Figure), and an IMF closer to a Kroupa-like distribution at larger distances.  For what concerns the quality of the fits to spectral indices, we show in App.~\ref{app:bestfitindices} that our models are able to match reasonably well all the observed line-strengths (for all galaxies, and radial bins).
Fig.~\ref{fig:IMF_methods} also shows that our results are independent of the fitting method.
In particular the \gammab\ profiles remain virtually indistinguishable  from the reference case
when considering BaSTI (rather than Padova00) stellar population models, or excluding Na features from the fits (see upper--left and lower--right panels in the Figure, corresponding to methods E and C, 
respectively). Negative IMF gradients are also found when excluding TiO features (method B; see lower--left panel) or when assuming 2SSP models (method D; see upper--right panel), although in these cases, the scatter is larger than in our reference fitting method. For method B, the scatter increases because excluding TiO features makes the fits more sensitive to the correlated variation of IMF slope and \nafe\ abundance ratio, while in method D, results are more sensitive to the degeneracy between age and IMF slope (due to the sensitivity of Balmer and TiO lines to both parameters; see LB13).

Figs.~\ref{fig:IMF_ref} and~\ref{fig:IMF_methods} show that the IMF radial profiles of all XSGs are very similar to each other, when plotted versus physical scale (i.e. galactocentric distance in units of kpc). XSG10 is the only object with a flatter trend at $\rm R \gtrsim 1$~kpc. The Pearson's correlation coefficient, $\rho$,  of \gammab\ versus R for the whole data-set (i.e. all galaxies and radial bins) indicates a very significant correlation, with a value of $\rho \sim  -0.9$ (see Fig.~\ref{fig:IMF_ref}).
On the contrary, as shown in App.~\ref{app:radii}, when plotting IMF profiles versus normalized galactocentric distances, $\rm R/R_e$, the profiles for different galaxies do not coincide, independent of the method used to measure the effective radius (e.g. S\'ersic vs. B+D fitting; see Fig.~\ref{fig:imf_slope_radii}).
Hence, a physical distance of $\sim 2$~kpc marks the characteristic scale of IMF radial variations in our sample of very massive ETGs.  Since other galaxy parameters (e.g. metallicity, and abundance ratios) do also change with galactocentric distance in ETGs, the fact that IMF slope correlates with galactocentric distance, does not imply, necessarily, that distance is the main  driver of IMF variations. We investigate this point further in the following sections.

Our results (Fig.~\ref{fig:IMF_ref}) are qualitatively consistent with other works reporting negative IMF radial gradients in (some) ETGs (e.g.~\citealt{NMN:15a, vanDokkum:2017, Sarzi:2018, Parikh:2018, Vaughan:2018})~\footnote{
 In contrast to these works, ~\citet{Alton:2017, Alton:2018}  measured the J-band luminosity-weighted light from dwarf stars in the IMF, $\rm f_{dwarf}$, finding very little (at the few-percent level) radial variations of $\rm f_{dwarf}$ in ETGs. We notice that this result is not inconsistent with our findings. In fact, because of our adopted (bimodal) IMF parametrization,  the trends in Fig.~\ref{fig:IMF_ref} imply radial gradients in $\rm f_{dwarf}$ at the 
level of a few percent (see the conversion between \gammab\ and $\rm f_{dwarf}$ in table~C2 of~\citealt{Alton:2017}).
}, as well as with our previous works focusing on radial trends for XSG1 only (LB16 and LB17). 
However, the present work explores, for the first time, IMF radial variations at the very high mass end of the galaxy distribution, for a representative, homogeneous sample of massive ETGs. Moreover, it is important to point out that other works have mostly focused on optical (or NIR) features only, while we combine features over a wide (optical+NIR) wavelength baseline. 

\begin{figure*}
 \begin{center}
\leavevmode
    \includegraphics[width=17cm]{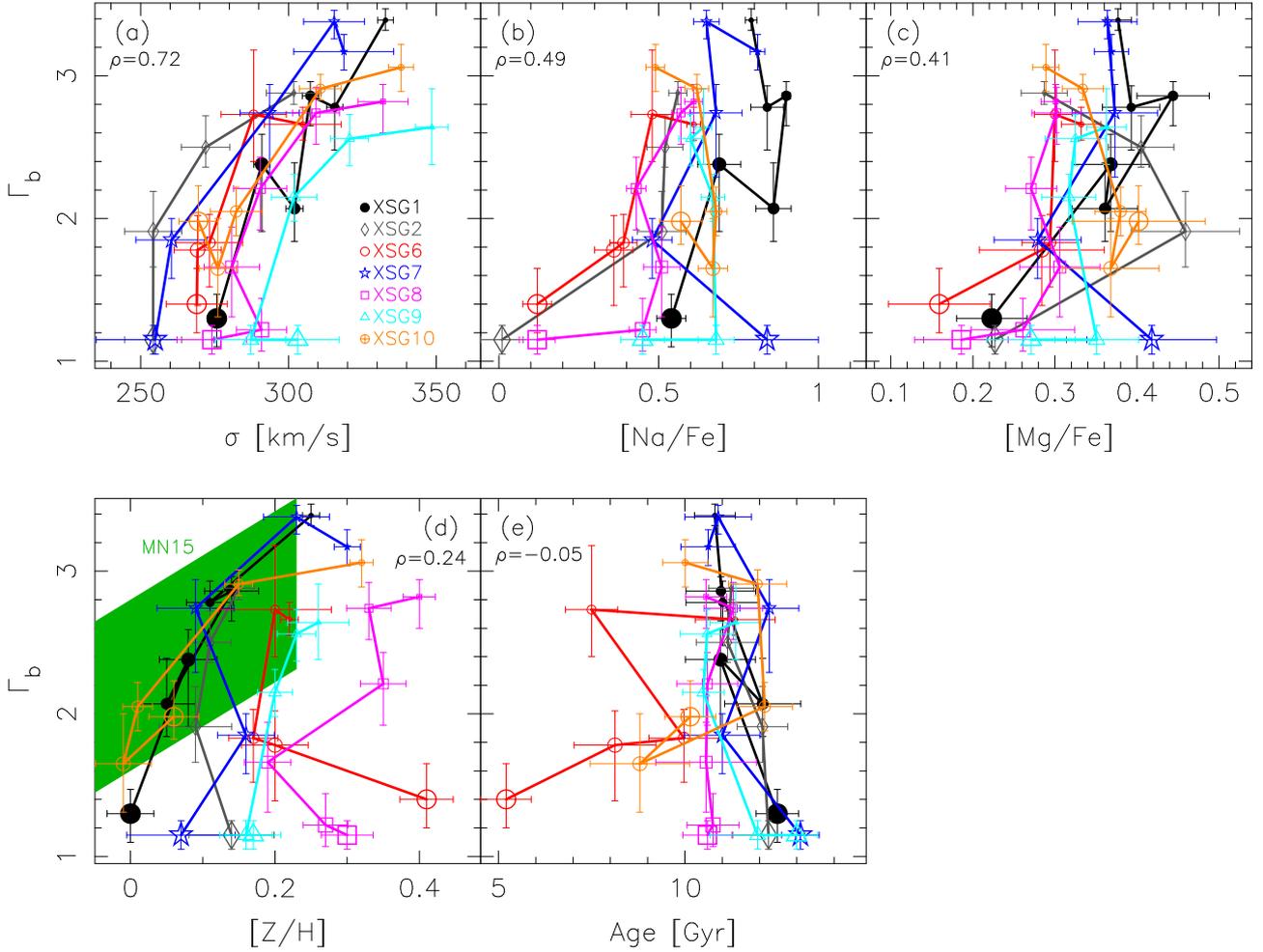}
 \end{center}
    \caption{
    The IMF slope, \gammab\,  is plotted against different quantities, namely velocity dispersion $\sigma$ (panel a),  Na and Mg abundance ratios (\nafe\ and \mgfe ; panels b and c, respectively), total metallicity \zh\ (d), and age (e). In each panel, we report the value of the Pearson's correlation coefficient, $\rho$, of \gammab\ versus each quantity, for all data-points (i.e. all galaxies/radial bins). The values of \gammab\ are for our reference method A (see Tab.~\ref{tab:methods}). 
     Symbols, line types, and error bars follow the same coding as in Figs.~\ref{fig:IMF_ref} and~\ref{fig:IMF_methods} (see also labels in the lower-right corner of panel a).     Notice that all correlations are weaker than that with galactocentric distance ($\rho \sim 0.9$; see Fig.~\ref{fig:IMF_ref}).
     The green shaded area in the bottom--left panel marks the locus of the IMF-\zh\ relation from MN15b (see the text).
    }
    \label{fig:gammab_gradients}
\end{figure*}

\subsection{Local drivers of IMF variations}
\label{subsec:IMFdrivers}

Fig.~\ref{fig:gammab_gradients} plots IMF slope, \gammab, for our reference fitting method versus different quantities (in different panels), i.e. (from top to bottom, and left to right) velocity dispersion ($\sigma$), \nafe\ and \mgfe\ abundance ratios, total metallicity \zh,  and age. 

\subsubsection{Radial trends of age, metallicity, and abundance ratios}
\label{subsubsec:gradients}

{ Radial trends of age, metallicity, and abundance ratios for our sample of ETGs (as plotted in Fig.~\ref{fig:gammab_gradients}), are generally consistent with previous studies of stellar population gradients in massive galaxies. The XSGs have rather flat age gradients, in agreement with, e.g., \citet[see references therein]{NMN2018}. The only exception is XSG6, whose age values tend to become younger outwards, consistent with the fact that this galaxy has the strongest ionization pattern among all XSGs, for all radial apertures (see App.~\ref{app:emission}).  
The \mgfe\ radial gradients are generally flat, with XSG1, XSG6, XSG8, XSG9 having lower \mgfe\ in the outermost bin, XSG2 and XSG7 having radially constant \mgfe\ profiles for all bins, and XSG10 showing a slightly positive \mgfe\ gradient. This is consistent with results of, e.g., ~\citealt{SanchezBlazquez:2007, Greene:2015, NMN2018, Parikh:2018}. In contrast, ~\citet{vanDokkum:2017} have reported positive \mgfe\ radial gradients for their sample of seven ETGs over a range of $\sigma$. 

For what concerns radial gradients of total metallicity, 
$\rm \nabla_Z= \delta [Z/H]/\delta(\log R)$, ~\citet{Spolaor:2009} 
reported a wide range of values at high galaxy mass, ranging from null gradients, to gradients  as steep  as -0.5~dex  per  decade  in  radius (see their figure~2).  Using  the best-fitting value  of $\rm [Z/H]$ from our XSG  spectra (as shown in panel  d of  Fig.~3), and performing an orthogonal least-square fit of $\rm [Z/H]$ vs. $\rm \log R$, we  obtain $\rm \nabla_Z=-0.26\pm 0.04$, $-0.10 \pm 0.05$, $-0.04 \pm 0.05$, $-0.23 \pm 0.06$, $-0.14 \pm 0.04$ , $-0.09 \pm 0.06$, $-0.41 \pm 0.04$, for XSG1, XSG2, XSG6, XSG7, XSG8, XSG9, and XSG10, respectively. These values are consistent with those reported by previous spectroscopic studies (see~\citealt{NMN2018} and references therein).

In the case of \nafe, most galaxies have rather flat gradients in the inner regions, that drop down significantly in the outermost bins for XSG2, XSG6, and XSG8.  On the other hand, XSG1~\footnote{Notice that \nafe\ and other stellar population properties for XSG1 are not exactly the same as those reported in LB16 and LB17, as in the present work we are fitting a different set of IMF-sensitive features with respect to those studies.} has a midly negative \nafe\ gradient, while XSG7, XSG9, and XSG10 tend to have flat gradients. Notice that we obtain Na abundance ratios as high as $0.8$--$0.9$~dex for some galaxies (XSG1 and XSG7) in our sample. Such high Na abundances in ETGs may reflect metallicity-dependent yields of Na from high- and intermediate-mass stars (see 
LB17 for a detailed discussion), though one should notice that, indeed, XSG1 and XSG7 are not the objects with the highest metallicity in our sample.
While other studies (e.g.~\citealt{vanDokkum:2017, Parikh:2018}) have reported significantly negative radial gradients of \nafe\ for ETGs, what matters for the present study is that our results remain virtually the same even when 
excluding Na features from the analysis (see Sec.~\ref{subsec:IMFgradients}). Moreover, we notice that the Na--MILES stellar population models adopted in the present work are the only set of models  taking the joint effect of \nafe\ and IMF explicitely into account (see LB17).
}

\subsubsection{Correlations with \gammab}

{ For each panel of Fig.~\ref{fig:gammab_gradients}, we report the Pearson's 
coefficient, $\rho$, of the correlation between \gammab\ and the corresponding 
parameter on the x-axis.}
Symbol size increases with galactocentric distance, R. Notice that each plot in Fig.~\ref{fig:gammab_gradients} can be compared directly with the trend of \gammab\ with galactocentric distance, as shown in Fig.~\ref{fig:IMF_ref}.   
Results can be summarized as follows:
\begin{itemize}
 \item[--] there is essentially no correlation ($\rho \sim 0$) of \gammab\ with age (see panel e of Fig.~\ref{fig:gammab_gradients}), both individually (i.e. for each single galaxy), and globally (for all data-points);
 \item[--] correlations with abundance ratios are mild, mainly due to points with lower \nafe\ and \mgfe\ in the outermost radial bins (see panels~b and~c of  Fig.~\ref{fig:gammab_gradients});
 \item[--] there is a significant correlation of \gammab\ with velocity dispersion ($\rho \sim 0.7$; see panel a of Fig.~\ref{fig:gammab_gradients}), with large scatter, in particular for $\sigma \lesssim 300$~\kms , where data-points populate a wide region with $1.2 \lesssim $\gammab$ \lesssim 2.5$;
 \item[--] individual galaxies show a correlation of \gammab\ with metallicity, but overall, we do not detect a global correlation of IMF slope with \zh, the correlation coefficient being $\rho \sim 0.24$; in particular, at \zh$\sim 0.2$ (where our results are not affected by any extrapolation of the models in the high metallicity regime, see Sec.~\ref{subsec:indices}), different spectra populate the whole available range of \gammab\ values, from a bottom-heavy slope (i.e. \gammab$\sim 3$, see the innermost data-points of XSG1 and XSG7), to a Kroupa-like distribution (\gammab$\sim 1.3$, see outermost data-points for XSG9 and XSG8); 
 \item[--] galactocentric distance is the quantity giving the tightest correlation to \gammab\ ($\rho =-0.91$; see Fig.~\ref{fig:IMF_ref} and Sec.~\ref{subsec:IMFgradients}) with respect to all other quantities in Fig.~\ref{fig:gammab_gradients}.
\end{itemize}

The lack of correlation of \gammab\ with \zh\ contrasts with the results of~MN15b, where the authors  analyzed ETGs in the CALIFA survey, finding that metallicity gives a better correlation to IMF variations with respect to dynamical (e.g. $\sigma$) and stellar population properties (e.g. \mgfe ; see also~\citealt{vanDokkum:2017}and~\citealt{Parikh:2018}). Panel d in Fig.~\ref{fig:gammab_gradients} further illustrates this point, comparing the trend of \gammab\ with \zh\ for the XSGs, with the locus of \gammab\ versus  \zh\ for CALIFA ETGs (from figure~2 of~MN15b), represented as a green shaded area. The XSGs show much steeper (almost vertical) tracks in the \gammab--\zh\ diagram with respect to  CALIFA ETGs, covering a wider range of \gammab's  at given \zh\ (see above). While this discrepancy might seem surprising, one should notice that the selection criteria of the XSG sample are remarkably different than those of ETGs in MN15b, as (i) the XSGs have been selected to provide a representative sample at the very high-mass end of the galaxy 
population (see Sec.~\ref{sec:specdata}); and (ii) 
none of the CALIFA ETGs in MN15b have central velocity dispersion larger than 300~\kms\ (see fig.~1 of MN15b), unlike the XSGs; (iii) as a consequence of point (ii), ETGs in CALIFA populate a \zh\ range from about -0.3 up to \zh$\sim +0.2$, while all XSG spectra are either in the solar or super-solar metallicity regime (see panel d of Fig.~\ref{fig:gammab_gradients}).  
Indeed, ETGs at very high-mass end might have different radial IMF gradients than the average population of ETGs -- depending on the way they formed and accreted their outer regions -- hence explaining the difference between our findings and those in MN15b (as well as some other studies of IMF radial gradients). We further discuss this point in Sec.~\ref{sec:Discussion}.

As shown in panel e of Fig.~\ref{fig:gammab_gradients}, we do not find any correlation of \gammab\ with age. Although we do not expect any significant contribution of young stellar populations in our sample of massive ETGs, we can investigate this point further, by looking for spectra with emission lines (if any) in the X-Shooter sample. As shown in App.~\ref{app:emission}, some of our galaxies, i.e. XSG1, XSG6, and XSG8,  show signs of emission lines in the optical spectral range. The analysis of diagnostic diagrams shows that  this emission implies a ionization pattern typical of retired stellar populations (i.e. emission lines produced by hot post-asymptotic giant branch stars) in the galaxy central regions (see the Appendix for detail). For these objects, one may expect that  1SSP and 2SSP models might not give an accurate description of the galaxy star-formation history, possibly affecting our results. However, as shown in Fig.~\ref{fig:gammab_gradients}, XSG1, XSG6, and XSG8 do not show different IMF profiles, neither they exhibit peculiar abundance patterns, compared to other galaxies. In other terms, the presence of (weak) emission in massive ETGs does not affect at all the IMF variations.

We also notice that the IMF profiles of the XSGs do not depend on  environment. Satellite galaxies in our sample (XSG2 and XSG7) have  IMF profiles similar to those for the rest of the sample, although XSG7 is the galaxy with bottom-heaviest IMF in the three innermost radial 
bins among all the XSGs (see Fig.~\ref{fig:IMF_ref}). However, as detailed in App.~\ref{app:envi}, the hierarchy of XSG7 is  uncertain; moreover, both XSG2 and XSG7 might have been actually centrals 2--3$ \, \rm Gyr$ before infalling into their current parent groups. In other words, all the XSGs might have spent most of their evolutionary history as group centrals. Notice that the lack of correlation between IMF profiles and galaxy environment is consistent with the results of~\citet{Rosani:2018}, who found that the IMF--$\sigma$ relation is independent of hierarchy.


\begin{figure}
\begin{center}
\leavevmode
    \includegraphics[width=7cm]{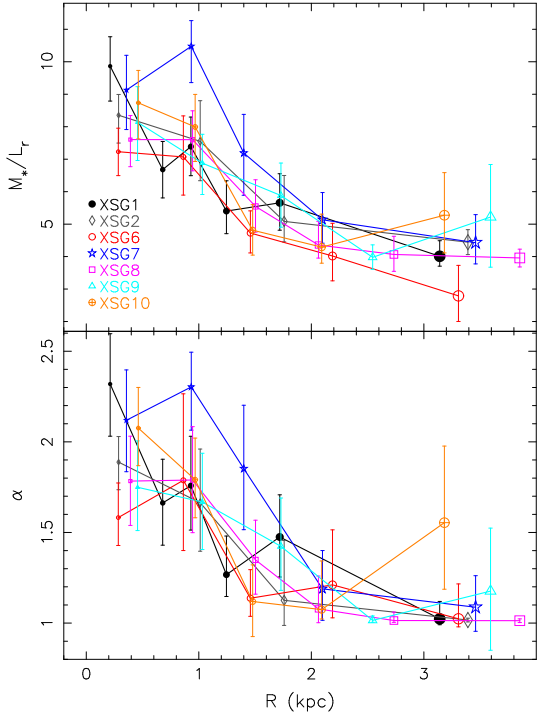}
 \end{center}
    \caption{Radial gradients of the r-band stellar mass-to-light ratio, \mlr\ (upper panel), and the ``mismatch parameter'' $\alpha$ (lower panel).
    The $\alpha$ is given by the estimated \mlr\ normalized to that for
    a Kroupa-like IMF (assuming the same age and metallicity).
    Error bars correspond to 1$\sigma$ uncertainties. In order to account for systematics in the \ml\ estimates, both \mlr\ and $\alpha$ are obtained by averaging out results from different methods (i.e. methods A, B, and E; see the text and Tab.~\ref{tab:methods}).
    Different galaxies are plotted with different symbols (whose size increases with galactocentric distance) and colours 
    (see labels in the upper panel), with the same coding as in Fig.~\ref{fig:IMF_ref}. 
    }
    \label{fig:ml_gradients}
\end{figure}

\subsection{$\rm M/L$ radial gradients}
\label{sec:ml_gradients}

Fig.~\ref{fig:IMF_ref} shows that all the XSGs exhibit a bottom-heavy IMF, i.e. an excess of low-mass stars, in the galaxy central regions.
Since low-mass stars dominate the mass budget of a stellar population, Fig.~\ref{fig:IMF_ref}  implies, in turn, a significant radial gradient in the stellar mass-to-light ratio. Fig.~\ref{fig:ml_gradients} (top panel) shows the expected r-band mass-to-light ratio, \mlr, for all the XSGs, as a function of galactocentric distance. Since the variation of IMF slope (with respect to that of other parameters, such as \zh ) dominates the \mlr , the top panel of Fig.~\ref{fig:ml_gradients} looks similar to Fig.~\ref{fig:IMF_ref}, i.e. all galaxies show a similar \mlr\ radial gradient,  with $6 \lesssim$\mlr$\lesssim 11$ in the center, down to \mlr $\sim 4$ in the outermost radial bins, the latter value corresponding, approximately, to the \ml\ expected for a Milky Way-like IMF. 

The \ml\ radial gradients are further shown in the bottom panel of Fig.~\ref{fig:IMF_ref}, where we plot the ``mass-excess'' parameter, $\rm \alpha$, in r band, as a function of galactocentric distance. The $\alpha$  is defined as the actual stellar mass-to-light ratio of a stellar population, normalized to that predicted for a Kroupa-like IMF, assuming  the same parameters (e.g. age and metallicity) as the given population. 
The $\alpha$ parameter has been first introduced by CvD12a to single out the effect of variations in the IMF with respect to that of  other parameters.
For a Kroupa-like IMF, one has $\alpha =1$.
In the innermost bins, the XSGs have $\alpha$ significantly above one, with values between $\sim 1.7$ (e.g. XSG6 and XSG9) and $\sim 2.2$ (e.g. XSG1 and XSG7), while at larger distances ($>2$~kpc), we find  $\alpha \sim 1$ (i.e. a Milky Way-like distribution), consistent with Fig.~\ref{fig:IMF_ref}. 

\begin{figure}
 \begin{center}
\leavevmode
    \includegraphics[width=7cm]{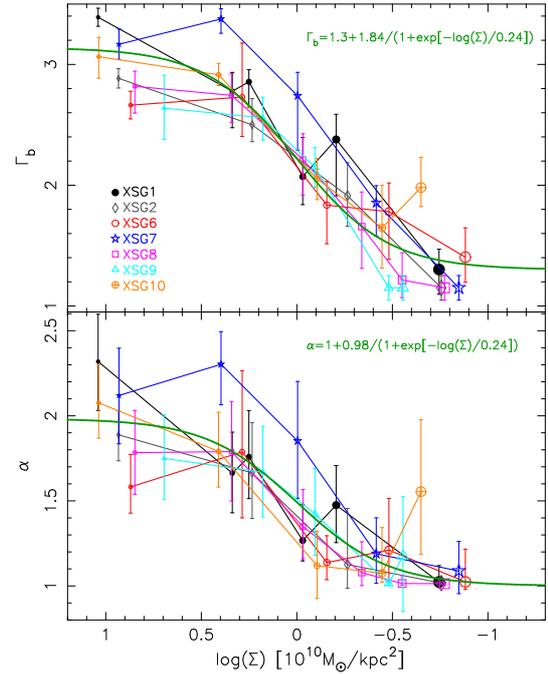}
 \end{center}
    \caption{IMF slope \gammab\ (top) and mismatch parameter $\alpha$ (bottom) are plotted as a function
    of logarithmic projected stellar mass density $\Sigma$, for our reference fitting scheme (method A; see Tab.~\ref{tab:methods}).
    Different galaxies are plotted with different symbols (with symbol size increasing with galactocentric distance) and colours 
    (see labels in the upper panel), with the same coding as in Fig.~\ref{fig:IMF_ref}. 
    Green curves are the best-fitting functional forms (whose equations are given in the upper--right of each panel) that describe the trend for all galaxies in our sample (see the text).
    }
    \label{fig:imf_dens}
\end{figure}

However, we warn the reader about a ``swift'' use of results in Fig.~\ref{fig:ml_gradients}. In fact, both the mass-excess factor and the stellar $\rm M/L$ depend significantly on the assumed IMF functional form (see LB13 and LB16). For instance, assuming a slightly higher low-mass end cutoff than that used in our models (i.e. $0.1 \, M_\odot$; see Sec.~\ref{sec:spmodels}), one could decrease the stellar M/L, without changing significantly the predicted line-strength of IMF--sensitive spectral indices (see, e.g., \citealt{Barnabe:2013}), as most features are actually insensitive to very low-mass stars~\footnote{An exception is the Wing-Ford band, \feh , at $\lambda \sim$9915\,\AA, whose sensitivity peaks in the very low-mass range, below $\sim 0.3 \, M_\odot$ (see CvD12a). As shown in LB16, the bimodal IMF is able to match the \feh\ measured in our spectra for XSG1, in contrast to a single power-law parametrization, predicting too high \feh . { However,  \feh\ is also strongly dependent on \mgfe\ and \nafe, and the FeH molecule is not implemented yet in the Na-MILES models (see LB17),} hampering the interpretation of this feature (see, e.g., \citealt{vanDokkum:2017, Vaughan:2018}). }. 
On the other hand, as shown in \citet{Lyubenova} (see also LB16), the (bimodal) IMF functional form provides consistent M/L estimates with respect to dynamical studies. We plan to come back to this issue in a forthcoming paper, comparing dynamical and stellar population M/L estimates for the XSGs.


The presence of steep \ml\ gradients in our sample of massive ETGs, implies that luminosity-weighted \ml\ ratios within a circular aperture depend significantly on the aperture size. In particular, within a region of 
1~\ret\ (the effective radius from the B+D decomposition, see Sec.~\ref{sec:spmodels}), our IMF profiles imply luminosity-weighted values of \alfa\ in the range from $1.15 \pm 0.05$ (for XSG8) to $1.6 \pm 0.1$ (for XSG7), with an average of $\alpha = 1.3 \pm 0.1$.  In other terms, under the IMF parametrization adopted in the present work, massive ETGs should have $\rm M/L$'s inside $\rm 1 \, R_e$, only $\sim 30\%$ larger on average with respect to the case of a radially constant Milky Way-like IMF. Notice that (i) to perform these estimates of $\alpha$, we have extrapolated the \ml\ profiles outside the radial range probed by X-Shooter, assuming that the IMF remains consistent with a Milky Way-like distribution ($\alpha=1$) outwards; and (ii) the luminosity-weighted \alfa\ within $\rm R_e$ would be even lower when considering effective radii from S\'ersic models (\res), as some galaxies (XSG6, XSG7, XSG8, XSG9) have \res\ significantly larger than \ret\ (Tab.~\ref{tab:strucpars}).

Our \ml\ estimates can be compared to those of \citet[hereafter SLC15]{SLC:2015}, who found three massive ETGs (named SNL-0, SNL-1, and SNL-2) having spectroscopic signatures consistent with a bottom-heavy IMF, but nevertheless a ``light'' \ml , as implied by lensing/dynamical constraints. Within a typical region of $\sim 2$~kpc, the SNL's have $\alpha = 1.06 \pm 0.1$ ($\alpha = 1.3 \pm 0.13$~\footnote{These values of \alfa,  and their errors, are computed by averaging values for SNL-0, SNL-1, and SNL-2 from table~1 of~\citet{Newman:2017}.}) in the case where dark-matter (no dark-matter) contribution, within the galaxy central regions, is taken into account (see table~1 of~\citealt{Newman:2017}). On the other hand, within a region of 2~kpc, for the XSGs, we estimate a median luminosity-weighted (circularized) \alfa\ of $\sim 1.5 \pm 0.15$, which is consistent with the SNLs only in the case of no dark-matter. Increasing the low-mass end cutoff in the IMF, would likely match better the \ml\ of SNL ETGs~\citep{Newman:2017}. However, this requires a somewhat fine-tuned scenario, whereby the IMF has an excess of low-mass stars (to match the spectroscopic signatures), but lacks the very low-mass stars (to match the \ml ).

\subsection{IMF variations and mass density}
\label{subsec:density}

Since all our targets have similar values of the central velocity dispersion (which is a proxy of galaxy mass, given the small  rotation velocity of the XSGs), the presence of IMF radial gradients in our sample  suggests 
a correlation of IMF slope with  mass density. Fig.~\ref{fig:imf_dens} shows the trend with surface mass density $\Sigma$, for both IMF slope, \gammab\ (top panel), and the ``mass-excess'' factor, $\alpha$, i.e. the mass-to-light ratio normalized to that for a Kroupa-like IMF (bottom-panel). Indeed, a significant correlation exists~\footnote{
Notice that the computation of $\Sigma$ involves the actual $M*/L$, that, in turn, depends on \gammab. 
In other terms, by construction, the variables on the x- and y-axes of Fig.~\ref{fig:imf_dens} are not independent. However, the relation in Fig.~\ref{fig:imf_dens} is a genuine one.
In fact, we found that the correlation of \gammab\ with luminosity (rather than mass) density is very similar to that of \gammab\ versus $\Sigma$.
}, that can be modeled with the analytic functions shown in Fig.~\ref{fig:imf_dens} (see green solid curves and equations), having a plateau at both high
 and low density (corresponding to a bottom-heavy and Kroupa-like IMF, respectively), and a transition region at $\rm \Sigma \sim 10^{10} M_\odot \, kpc^{-2}$. This characteristic density  corresponds to a characteristic scale of $\sim 1$--$2$~kpc, as in Fig.~\ref{fig:IMF_ref}. { However, we caution the readers that the high-density plateau in Fig.~\ref{fig:imf_dens} is (partly) driven by the IMF determination for the central radial bins of the XSGs, where some extrapolation of stellar population models to the high-metallicity regime is required (see Sec.~\ref{sec:spmodels}).}

The correlation in Fig.~\ref{fig:imf_dens} may appear in disagreement with the result of~\citet[hereafter SBK15]{SpiBar:2015}, who claimed an anticorrelation of IMF slope and galaxy mass density. We notice, however, that the finding of SBK15 refers to the IMF slope in the central galaxy regions versus total mass density (the latter estimated as $\rm \sigma^2 / R_e^2$), while in Fig.~\ref{fig:imf_dens} we show a {\it local} correlation of IMF slope and (stellar) mass density. Since all the XSGs have a similar $\rm \sigma_0$ (especially when considering the central circular aperture probed by the SDSS fiber, see Tab.~\ref{tab:galaxies}), ``total'' mass density in our sample is inversely proportional to $\rm R_e^2$. Indeed, although the galaxies with largest  $\rm R_e$ in our sample (i.e. XSG6, XSG8, and XSG9) tend to have a lower IMF in the central radial bins (Fig.~\ref{fig:IMF_ref}) than the others, the IMF in their center is still bottom-heavy  (\gammab$\sim 2.75$~\footnote{This is the average of \gammab\ values at $R \sim 0$ for XSG6, XSG8, and XSG9, when considering our reference fitting scheme.}), and overall, their IMF radial gradients are similar to those for the other XSGs~\footnote{In fact, computing the luminosity-weighted value of $\alpha$ for XSG6, XSG8, and XSG9, within a mock circular aperture of radius 1'', gives $1.74 \pm 0.04$, which is fully consistent with the average value of $1.94 \pm 0.2$ for the other XSGs.} (Fig.~\ref{fig:IMF_ref}). Therefore, our data do not point to a fundamental anticorrelation of IMF slope and total density (as in SBK15), but rather, a significant correlation of local IMF slope and projected mass density. 

{ Interestingly, ~\citet{Barone:2018} have recently found a correlation of stellar 
population properties, such as age and \mgfe , with surface mass density, interpreting these
trends as a result of compactness-driven  quenching  mechanisms, and/or as a consequence 
of star-formation rate (SFR)--gas density relation in the disk-dominated progenitors of ETGs. 
We notice that our sample of ETGs does not exhibit significant age and \mgfe\ radial gradients (see Sec.~\ref{subsubsec:gradients}), but nevertheless the IMF slope correlates with surface-mass density. Therefore, IMF variations in our sample do not seem to have necessarely the same origin as the stellar population trends found by ~\citet{Barone:2018}. 
}

\section{Discussion}
\label{sec:Discussion}

All massive ETGs  in our sample (XSGs) show very similar, steep, IMF radial gradients, over a physical region of $\sim 4$~kpc from the galaxy centers. The IMF changes from bottom-heavy, to Kroupa-like, at a characteristic distance of $\sim 2$~kpc (Fig.~\ref{fig:IMF_ref}), or, alternatively, at a typical surface mass density of $\sim 10^{10} \, M_\odot \, kpc^{-2}$ (Fig.~\ref{fig:imf_dens}). 
In a two-phase model of galaxy formation, the core of massive galaxies is expected to form ``in-situ'' at high redshift through a rapid dissipative process of star formation, while the outer galaxy envelopes are expected to build up at later epochs through major/minor mergers~\citep{Oser:10, Oser:2012, NavarroGonzalez:2013}. This might be the case also for massive BCGs~\citep{Laporte:2013, Oogi:2017} -- an aspect that is relevant for the present work, as most of our targets are actually group centrals, or might have been centrals before accretion onto a galaxy group a few Gyrs before the epoch of observation (see App.~\ref{app:envi}). The characteristic scale of $\sim 2$~kpc for IMF gradients in our sample coincides with the typical size of high-redshift massive compact galaxies (the so-called ``red nuggets''; see, e.g.,\citealt{Trujillo:2007, Damjanov:2011, Saracco:2011}). This suggests that our IMF gradients are connected to those of massive compact galaxies. Indeed, a few massive compact galaxies have been found also at z$\sim$0, the prototype of these objects being NGC\,1277~\citep{Trujillo:2014}. These galaxies, that are believed to be the ``relics'' of the high-z red nuggets, have significant rotation and velocity dispersion gradients, with shallow IMF radial gradients (the IMF being bottom-heavy) over a region of at least 3--4~kpc~\citep{NMN:15c, Anneta:2017}. More in detail, \citet{Anneta:2017} found that at a galactocentric distance of $R \sim 3$~kpc (corresponding to a mass density of $\rm \sim 10^9 M_\odot/kpc^2$), the IMF slope of relic galaxies is \gammab$\gtrsim 2.5$  (see their figure~5), which is inconsistent with the trends  in Figs.~\ref{fig:IMF_ref} and~\ref{fig:imf_dens}. Moreover, in~\citet{NMN2018}, we found that relic galaxies tend to have \mgfe\ profiles that increase outwards, while normal ETGs, as well as the XSGs (see Fig.~\ref{fig:gammab_gradients}), tend to have either flat, or negative, \mgfe\ radial gradients.
Therefore, either the central regions of massive ($\sigma > 300$~\kms ) ETGs formed ``in-situ'' by a different channel than massive relic galaxies at $z \sim 0$, or other processes, such as merging/accretion affected significantly the region from $\sim 2$ to $4$~kpc, producing the steep IMF gradients we observe in our sample. 
Dry merging would unlikely explain, by itself, our observations. In~MN15b, we found a tight correlation of local metallicity and IMF slope for intermediate-mass ETGs, as probed by the CALIFA survey (see also~\citealt{Parikh:2018}). Therefore, any accreted material in massive ETGs would likely follow the CALIFA IMF--\zh\ relation, while as noted in Sec.~\ref{subsec:IMFdrivers}, at fixed metallicity we find a wide range of IMF slopes in our sample. In particular, at super-solar metallicity (\zh$\sim 0.2$), we find IMF slopes ranging from a bottom-heavy to a Kroupa-like distribution among different XSGs. As discussed in~\citet{Weidner2013} and~\citet{Ferreras2015}, in order to reconcile an excess of low-mass stars with the high metallicity content of massive galaxies, one may invoke a scenario whereby star formation starts with a top-heavy phase, switching to a bottom-heavy distribution at later times. Our results imply that the regions  corresponding to a few kpc in very massive galaxies may have formed/evolved through dissipative processes (e.g. wet mergers at high z), whose physical conditions were different than those in high-redshift compact cores, hence preventing an efficient switch from a top to bottom-heavy phase at all radii (see~\citealt{Vazdekis1997}). 

We point out that the present work is the first attempt to perform a systematic study of IMF variations in very massive 
($\sigma > 300$~\kms ) ETGs. 
{ ~\citet{Sarzi:2018} derived} the IMF radial profile 
for M87 (the giant elliptical galaxy at the center of the Virgo cluster), finding a consistent profile to that for XSG1. 
\citet{Vaughan:2018} also derived spatially  resolved  measurements  of the  stellar  IMF in NGC\,1399, the largest elliptical galaxy in the Fornax Cluster. They found a super-Salpeter IMF out to $\rm \sim 0.7 \, R_e$, with a distribution marginally  consistent  to a  Milky--Way  IMF  beyond $\rm R_e$, corresponding to a characteristic scale of $\sim 3$--$4$~kpc. This scale is significantly larger than that for the XSGs and M87, meaning that a wider range of IMF profiles might exist, than those for our XSG sample. This might also be the case for SNL-0~\citep{SLC:2015, Newman:2017}, a very massive ($\sigma > 300$~\kms ) ETG whose lensing-based IMF normalization is consistent with a Milky-Way  distribution within a region of $\sim 1.9$~kpc. We notice that M87 has a rather flat velocity dispersion profile, with little rotation (less than $\sim$20~\kms\ within the region probed by~\citealt{Sarzi:2018}), as it is also the case for the XSGs. On the contrary, NGC\,1399, as well as compact massive ETGs, have significant velocity dispersion and rotation gradients. Therefore, we speculate that there might exist some {\it anti-correlation} of IMF and kinematics radial gradients, with galaxies showing shallower kinematics profiles having steeper IMF gradients. Qualitatively, such anticorrelation might result from the general properties of the Jeans mass, $\rm M_J \propto T^{3/2} \rho^{-1/2}$, with $\rm T$ and $\rho$ being  temperature and density, respectively. Assuming that local velocity dispersion ($\sigma$) is a proxy for $\rm T$, and given the shallow $\sigma$ gradients of the XSGs, one would have $\rm M_J \propto \rho^{-1/2}$, i.e. the correlation of IMF and mass density in Fig.~\ref{fig:imf_dens} would be a consequence of 
the dependence of $\rm M_J$ on local density. On the contrary, in systems with a stronger $\sigma$ drop with radius (such as the relic galaxies), the decrease of both $\rm T$ and $\rho$ with radius would imply a more constant $\rm M_J$, i.e. shallower IMF radial gradients (as observed). 

Overall, our results imply that there seems to be no single driver of IMF variations in early-type galaxies, and in particular, that metallicity cannot be the only culprit of IMF variations in stellar systems. A similar conclusion has been drawn by~\citet{Villaume:2017}, who analyzed IMF variations in compact, low velocity dispersion, stellar systems within a wide metallicity range, finding smaller IMF variations than in ETGs. \citet{NMN2019} also found that the two-dimensional map of IMF variations in a lenticular galaxy does not resemble  its metallicity (2D) distribution. 
Instead, in our sample of massive ETGs, we find a clear correlation of IMF slope with local mass density, as well as galactocentric distance in physical units.
These results, together with previous findings on IMF radial gradients, show that IMF variations in galaxies likely result from the complex formation processes and mass accretion history of galaxies at different galactocentric distances, showing the importance of determining IMF gradients to constrain the galaxy formation scenario.




\section*{Acknowledgements}
Based on observations made with ESO Telescopes at the Paranal
Observatory under programmes 
ID 092.B-0378, 094.B-0747, 097.B-0229 (PI: FLB).
{ We thank the anonymous referee for the helpful comments and suggestions.}
FLB acknowledges the Instituto de Astrof\'isica de Canarias for the
kind hospitality when this project started.  We thank dr.  J. Alcal\'a
for the insightful discussions and help with the reduction of
X-Shooter spectra. We have made extensive use of the SDSS database
(http://www.sdss.org/collaboration/credits.html).  FLB, AV, JFB
acknowledge support from { grant AYA2016-77237-C3-1-P} from the Spanish
Ministry of Economy and Competitiveness (MINECO). GvdV acknowledges funding from the European Research Council (ERC) under the European Union's Horizon 2020 research and innovation programme under grant agreement No 724857 (Consolidator Grant ArcheoDyn).
FLB also acknowledges  financial  support from  the European  Union's Horizon  2020  research  and  innovation  programme  under  the  Marie Sklodowska-Curie grant agreement n.721463 to the SUNDIAL ITN network.








\appendix

\section{The environment of XSGs}
\label{app:envi}
In order to assess the host environment of our targets, we matched the list of XSGs
with the group catalogue of~\citet{Wang:2014}, obtained from 
SDSS Data Release 7 (DR7), using the procedure of~\citet{Yang:2007}. In short,
the adaptive halo-based group finder of~\citet{Yang:2007} is applied to 
all galaxies in the Main Galaxy Sample of the New York University
Value-Added Galaxy Catalogue~\citep{Blanton:2005} for DR7~\citep{Abaz:2009}, 
in the redshift interval 0.01 $\leq z \leq$ 0.20, with a redshift completeness 
C$_z >$ 0.7, and an extinction corrected apparent magnitude brighter than $r = $ 
17.72 mag. Galaxy stellar masses are computed using the relations between stellar 
mass-to-light ratio and colour from~\citet{Bell:2003}.
The algorithm of~\citet{Yang:2007} is based on the traditional FOF method with small 
linking lengths in redshift space to arrange galaxies into groups and estimate the group 
characteristic stellar mass and luminosity. In the first iteration, the adaptive halo-based 
group finder uses a constant mass-to-light ratio of 500 $h$M$_{\odot}$/L$_{\odot}$ to 
evaluate a tentative halo mass for each group. This mass is later used to estimate the 
size and velocity dispersion of the group halo, which in turn define group membership 
in redshift space. Next, a new iteration starts, in which the group characteristic 
luminosity and stellar mass are converted into halo mass through the halo occupation 
model of~\citet{Yang:2005}. This procedure is reiterated until no more changes arise
in  group membership. Additional outputs of the algorithm are the group virial radius,
the satellite projected distance from the luminosity-weighted centre of the group, and 
the distinction of group members between centrals (the galaxies with largest
stellar mass) and satellites. 
\par\noindent
From the above mentioned group catalogue, we retrieved information about hierarchy and host
halo mass for our targets (see Tab.~A1). Most of the XSGs, i.e. XSG1, XSG6, XSG8, XSG9, and XSG10 are classified 
as central galaxies, while XSG2 and XSG7 are satellites, with a projected distance of 0.45R$_{\rm vir}$ and 0.08R$_{\rm vir}$ from the luminosity-weighted centre of their parent group, respectively. For what concerns XSG7, in practice this system is hosted by a group with two bright galaxies at its center, one of them being XSG7 and the other being a galaxy with similar luminosity (classified as group central). Therefore, although classified as satellite, the nature of XSG7 is uncertain.
Having stellar masses very similar to those of the brightest group galaxies of their parent groups (with stellar mass differences $\lesssim 0.15$dex), both XSG2 ans XSG7 might have  been centrals of recently infallen groups. To test this hypothesis, we also analyzed the phase-space of galaxies in the parent groups of XSG2 and XSG7 (using the same approach as in ~\citealt{Pasquali:2019}), finding no clear evidence of multiple components. Notice that, as shown by the numerical simulations of~\citet{Vijayaraghavan:2015}, accreted groups may be not detectable in phase space a few Gyr after infall, implying that both XSG2 and XSG7 could still have been centrals for most of their evolutionary history.
In such case, all radial profiles (and in particular IMF gradients) of the XSGs would be representative
of the population of central  ETGs at high-mass end.
For what concerns parent halo mass, XSG1, XSG2, XSG8, and XSG9, reside 
in hosts with a halo mass between 10$^{14}$ and 10$^{14.5}$
M$_{\odot}~h^{-1}$, whereas XSG6, XSG7 and XSG10 belong to environments in the range
10$^{13.5}$ - 10$^{14}$ M$_{\odot}~h^{-1}$. 

\begin{table}
\centering
\small
 \caption{Environmental properties of  XSGs. Column~1 gives the galaxy number in our sample, 
 column~2 reports if the galaxy is classified as a central (CEN) or satellite (SAT) in the group catalogue (see the text), while column~3 reports the halo mass of the parent halo where the galaxy resides.
 }
\begin{tabular}{c|c|c}
   \hline
  XSG\# & hierarchy & $\rm log10(M_h)$ \\
       &           &   ($M_\odot$) \\
   (1) &  (2) &   (3) \\
  \hline
 1 & CEN & 14.00 \\
 2 & SAT & 14.35 \\
 6 & CEN & 13.85 \\
 7 & SAT & 13.78 \\
 8 & CEN & 14.33 \\
 9 & CEN & 14.45 \\
10 & CEN & 13.52 \\
  \hline
  \end{tabular}
\label{tab:envi}
\end{table}

\section{Comparison of SDSS and X-Shooter central spectra}
\label{app:SDSS}

\begin{figure}
 \begin{center}
\leavevmode
   \includegraphics[width=9cm]{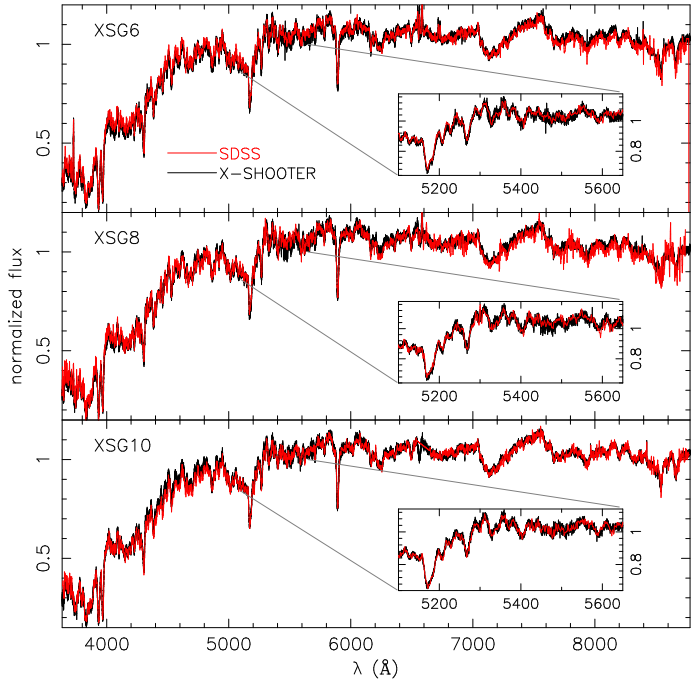}
 \end{center}
   \caption{Examples of X-Shooter (black) and { SDSS (red) optical spectra}, for  three galaxies in our sample. The X-Shooter spectra have been extracted to mimic the same circular aperture as the SDSS fiber (see the text). A very good agreement is found, even in the region from 5200 to 5600~\AA\ restframe (see the inset panels), which is corrected for the effect of the X-Shooter dichroic (see the text). 
   }
    \label{fig:comp_SDSS_XSHOOTER}
\end{figure}

All galaxies in our sample have SDSS spectroscopy available, in the optical spectral range, covering a fiber circular aperture of radius 1.5''. Fig.~\ref{fig:comp_SDSS_XSHOOTER} shows a representative comparison of X-Shooter and SDSS spectra for three galaxies in our sample (XSG6, XSG8, and XSG10). In order to mimic the SDSS circular aperture, the X-Shooter long-slit spectra have been extracted by weighting each point along the slit with a factor $\pi R$, where $R$ is the distance from the photometric center of the galaxy. Both X-Shooter and SDSS spectra have been corrected for Galactic reddening, with wavelengths given at restframe, in the air system. Fig.~\ref{fig:comp_SDSS_XSHOOTER} shows an excellent agreement (at few percent level) between the SDSS and X-Shooter spectra, despite the fact that we are comparing fiber to long-slit data. As expected (because of the higher spectral resolution), the X-Shooter spectra have far better quality, than the SDSS ones, in the regions strongly affected by telluric/airglow sky lines (e.g. in the CaT region, at $\lambda  \! \sim \! 8600$~\AA). The insets in the Figure zoom out the wavelength window, where the UVB and VIS X-Shooter spectra have been merged during the reduction process, at $\lambda \! \sim \! 5400$~\AA\ restframe (see LB16). This region is potentially affected by the temporal variation of the instrumental throughput due to the UVB/VIS dichroic element (see, e.g.,~\citealt{SCH:2014}). The comparison of SDSS and X-Shooter spectra shows a very good agreement also in the dichroic region. In particular,  we verified that the line-strengths of Fe5270 and Fe5335 spectral features (which enter our analysis, and fall in the dichroic region) are fully consistent (at 1 sigma level) between the two data-sets, further confirming the reliability of our data reduction procedure.

\section{Surface Photometry}
\label{sec:surfphotometry}

\begin{figure*}
 \begin{center}
\leavevmode
   \includegraphics[width=8.5cm]{f7.ps}
   \hspace{0.1cm}
   \includegraphics[width=8.5cm]{f8.ps}
   \includegraphics[width=8.5cm]{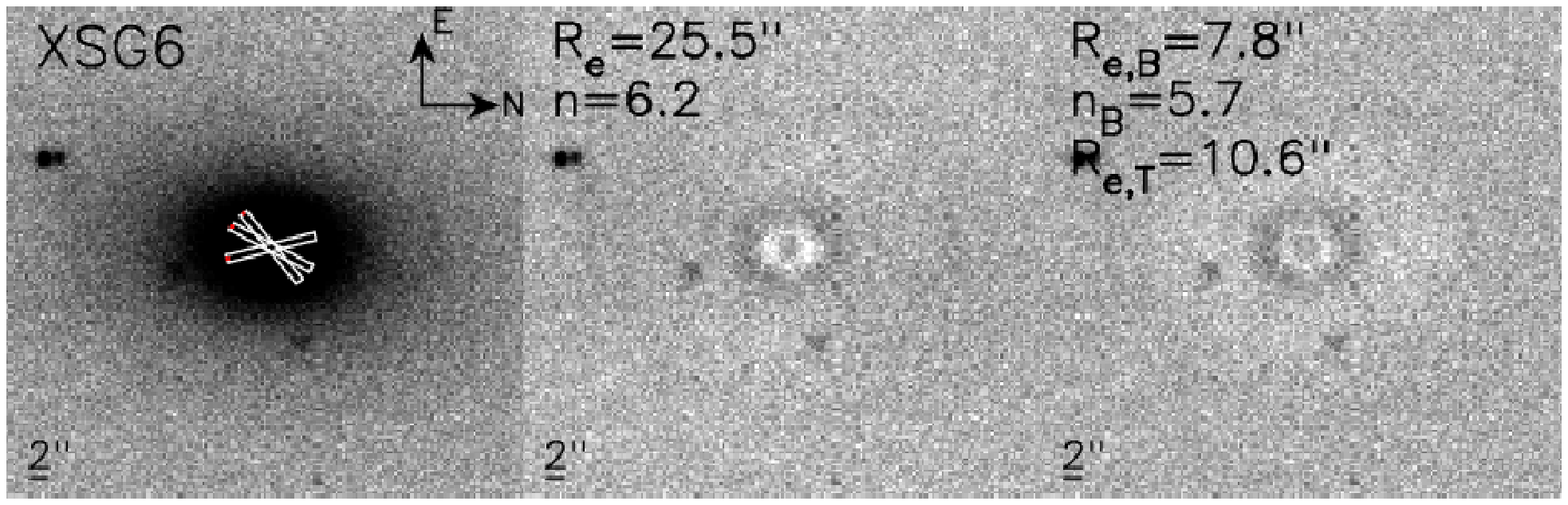}
   \hspace{0.1cm}
   \includegraphics[width=8.5cm]{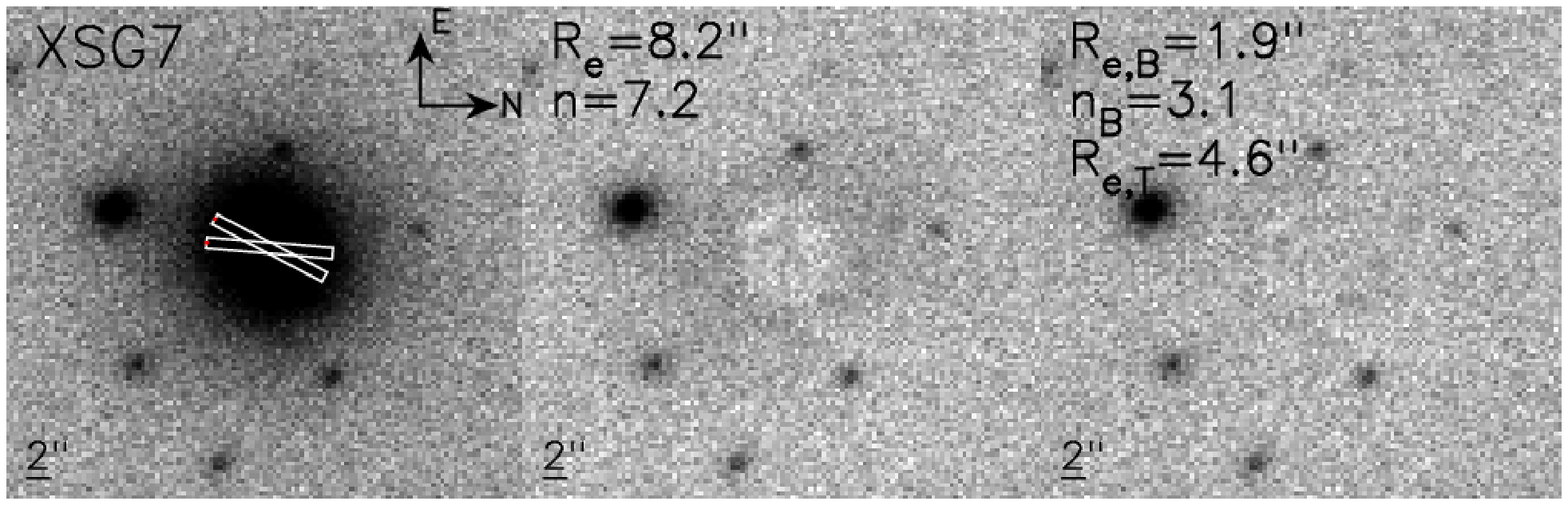}
   \includegraphics[width=8.5cm]{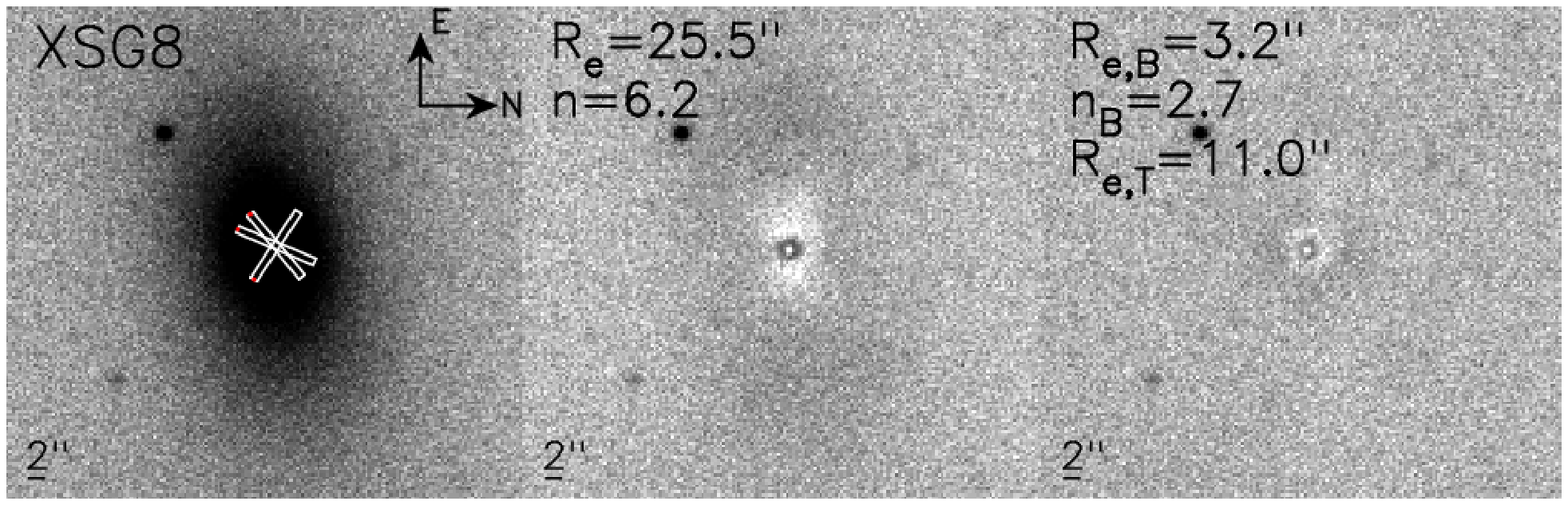}
   \hspace{0.1cm}
   \includegraphics[width=8.5cm]{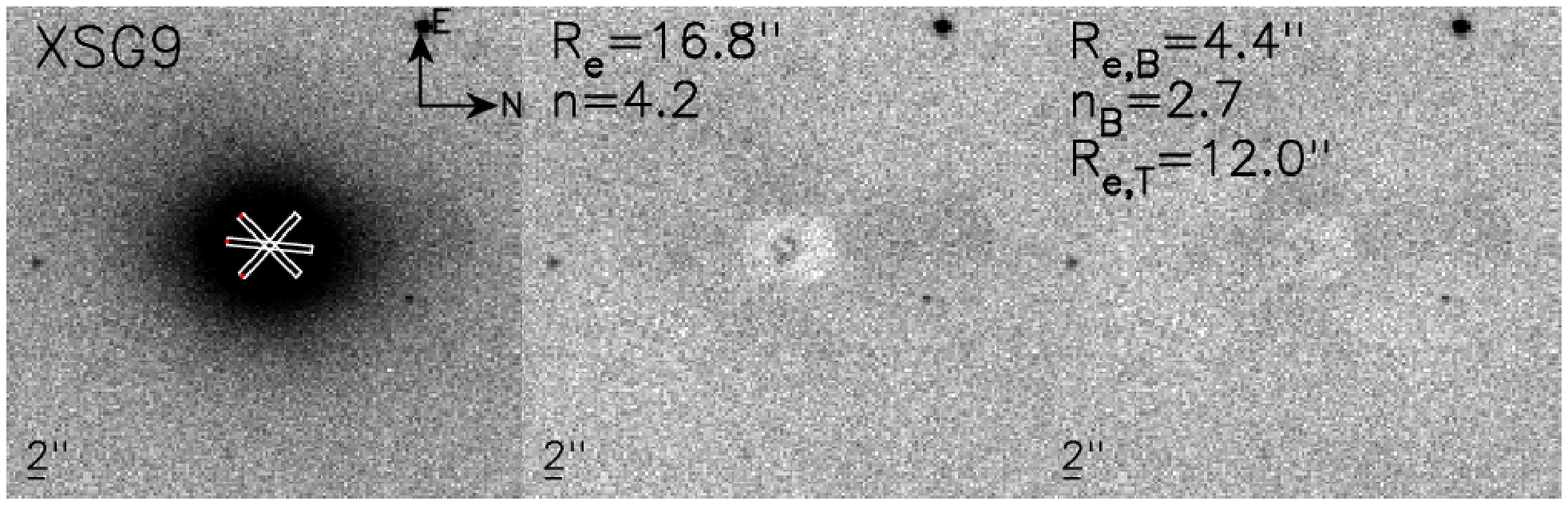}
   \vspace{0.5cm}
   \includegraphics[width=8.5cm]{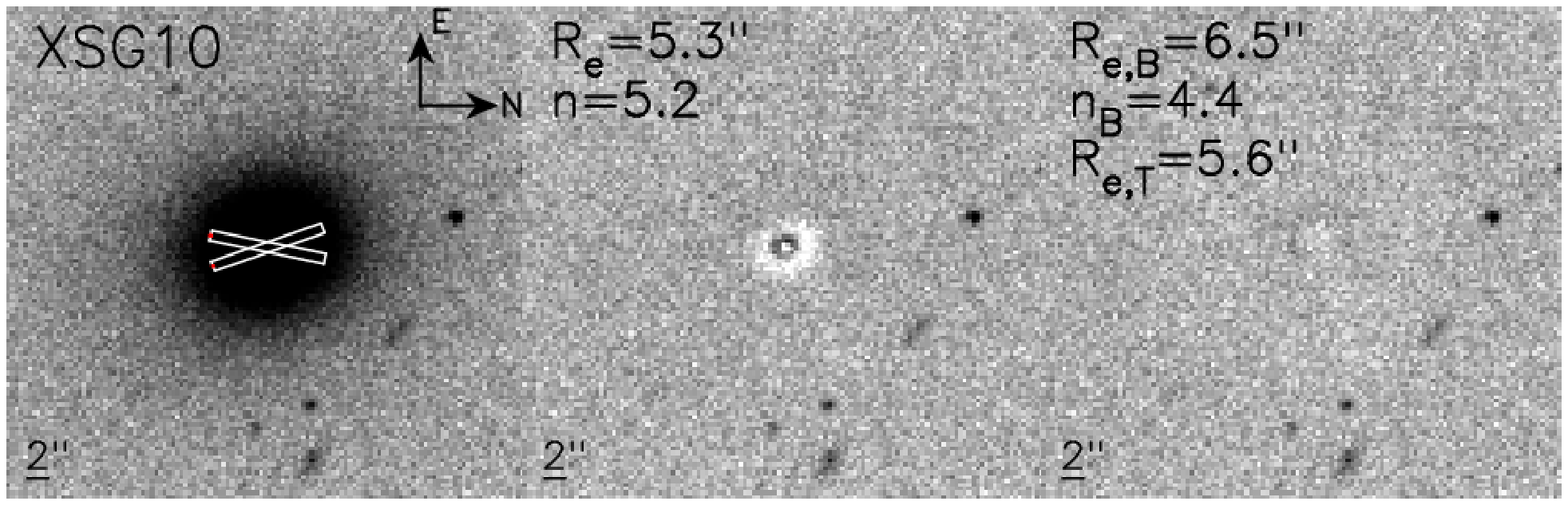}
 \end{center}
   \caption{Two-dimensional surface brightness decomposition of the SDSS r-band images of our massive galaxies targeted
   with X-Shooter (see Tab.~\ref{tab:galaxies}). From left to right, and top to bottom, galaxies are shown in the same order as listed in Tab.~\ref{tab:galaxies}.
    For each galaxy, the Figure plots three horizontal panels, showing the r-band SDSS galaxy image (left), with the galaxy ID on the top-left corner; the residual map, normalized to the expected noise in each pixel, after  subtracting the best-fitting Sersic model (middle);  the same as in the middle panel, but for the (Sersic-)bulge plus disc (B+D) decomposition (right). In the top--left corner
    of each middle  sub-panel, we report the effective radius, \res , and the shape parameter, \ns , of the best-fitting Sersic models. In the right sub-panels, we report the effective radius, \reb , and the shape parameter, \nb , of the best-fitting bulge component, as well as the half-light radius of the B+D model, \ret . The relevant best-fitting parameters are summarized in Tab.~\ref{tab:strucpars}.
   }
    \label{fig:2dfits}
\end{figure*}

In order to characterize the light distribution of the XSGs, and in particular the scale radius of their light profiles, we have used the software 2DPHOT~\citep{LBdC08} to 
fit the SDSS r-band images of all galaxies with two-dimensional, PSF-convolved, models. 
We have considered (i) single component S\'ersic models, and (ii) models with a S\'ersic bulge, 
plus an exponential (disc) component (hereafter B+D). Galaxy images, as well as residual maps 
after subtracting the best-fitting models, are shown in Fig.~\ref{fig:2dfits}.
Relevant parameters of the fits, namely the effective radius of the single S\'ersic and B+D fits, \res\ and \ret~\footnote{The \ret\ is computed by constructing a growth light curve from the best-fitting B+D model, and deriving the radius enclosing half of the total light of the model.}, respectively,
the axis ratio of the single-component fits, $\rm b/a$, the total magnitudes of the S\'ersic and B+D 
models, \mts\ and \mtbd , respectively, and the total light fraction in the bulge, $\rm B/(B+D)$, are reported in 
Tab.~\ref{tab:strucpars}. 

Fig.~\ref{fig:2dfits} shows that while, in general,  a single S\'ersic component describes reasonably well the 
XSGs' light distribution, the S\'ersic fits are not perfect, leaving either small central residuals
(see, e.g., the case of XSG10) or large-scale faint features in the model-subtracted images (see, e.g., XSG8).
Most of the residual features disappear when adopting a second component, through a B+D decomposition, 
with the exception of XSG6, for which both S\'ersic and B+D fits give similar residual maps. 
Notice also that the second component is significant, accounting for a 
fraction $\gtrsim 50 \%$ of the total galaxy light in most systems (XSG1, XSG2, XSG7, XSG8, and XSG9). We point out here that the lack of significant rotation in all the XSG's indicates that the second component describes actually an halo envelope, rather than a true (kinematical) disc. 
This result is consistent with the fact that most of XSGs are central group galaxies (see App.~\ref{app:envi}), that may have built up their extended envelopes through several minor mergers/smooth accretion. We also find that most of the XSGs are round objects, with an axis ratio $\sim 0.8$--$0.9$ (i.e. an E1--E2 morphological type), while XSG2 and XSG8  have a flatter shape ($\rm b/a \sim 0.6$), corresponding to an E4 type.

Tab.~\ref{tab:strucpars} shows that for more than half of the XSGs the effective radius 
depends significantly on the fitting method. In fact, while for XSG1, XSG2, and XSG10, the difference
between \res\ and \ret\ is within $\sim 20 \%$, for XSG6, XSG7, XSG8, and XSG9, the \res\ is 
significantly larger, by a factor about two, than \ret ~\footnote{Notice that XSG1 and XSG10 are the only two galaxies with \res $<$ \ret . For XSG10, this is due to the fact that the B+D model  consists of an extended bulge component (with \reb 
$\sim 6.5$~kpc$>$~\res ) and a very small, faint, disk  
($ \rm R_{e,D} \sim 0.5$~kpc). 
Hence, the total \ret\  is driven by \reb . For XSG1, the disk component is
significant, and more extended ($\rm R_{e,D} \sim 10$~kpc) than the bulge (\reb $\rm \sim 1.8$~kpc), making \ret\ larger than \res .
}.
The effective radius of the bulge component does also show a wide range of values, from $\sim 2$ to $\sim 8$~kpc, with a characteristic (mean) value of $\sim 3.9$~kpc.  Notice that this characteristic scale matches the radial range covered by our radially binned X-Shooter spectra.

\begin{table*}
\centering
\small
 \caption{Structural parameters of XSG galaxies in r band. Column~1 is the XSG label.
 Columns~2, 3, 4, and 5, report the best-fitting Sersic parameters, namely the effective radius, \res, the axis ratio \ba, the shape parameter \ns , and total magnitude \mts. Columns~6, 7, 8, 9, 10 provide the most relevant best-fitting parameters of the B+D decomposition, i.e. the effective radius and Sersic index of the bulge component, \res\ and \nb, the total (i.e. B+D) effective radius, \ret, the bulge luminosity fraction, \bt, as well as the total magnitude, \mtbd, respectively.  }
  \begin{tabular}{c|c|c|c|c|c|c|c|c|c}
   \hline
XSG\# & \res  &  \ba & \ns & \mts  &  \reb  & \nb   & \ret  & \bt   & \mtbd \\
      & (kpc) &      &     &  (mag) & (kpc) &       &  (kpc)  &    &  (mag) \\
 (1)  &  (2)  &  (3) & (4) &   (5) &  (6) &  (7)  &  (8)  &  (9)  &  (10) \\
  \hline
$ 1$  & $ 4.1$ & $0.90$ & $4.02$ & $14.40$ & $1.8$ & $2.78$ & $4.84$ & $0.51$ & $14.43$ \\
$ 2$  & $ 4.2$ & $0.63$ & $4.58$ & $14.61$ & $1.6$ & $3.51$ & $3.95$ & $0.54$ & $15.09$ \\
$ 6$  & $20.4$ & $0.75$ & $6.81$ & $13.32$ & $7.8$ & $5.73$ & $10.6$ & $0.70$ & $13.83$ \\
$ 7$  & $ 8.2$ & $0.84$ & $7.15$ & $14.25$ & $1.9$ & $3.14$ & $4.6$  & $0.55$ & $14.45$ \\
$ 8$  & $25.5$ & $0.63$ & $6.15$ & $12.98$ & $3.2$ & $2.66$ & $11.0$ & $0.39$ & $13.25$ \\
$ 9$  & $16.8$ & $0.84$ & $4.16$ & $13.22$ & $4.4$ & $2.66$ & $12.0$ & $0.40$ & $13.52$ \\
$10$  & $ 5.3$ & $0.85$ & $5.20$ & $14.17$ &$ 6.5$ & $4.43$ & $5.63$ & $0.93$ & $14.25$ \\
\hline
  \end{tabular}
\label{tab:strucpars}
\end{table*}

\section{IMF profiles versus  $\rm R/R_e$.}
\label{app:radii}

\begin{figure}
 \begin{center}
\leavevmode
    \includegraphics[width=8.5cm]{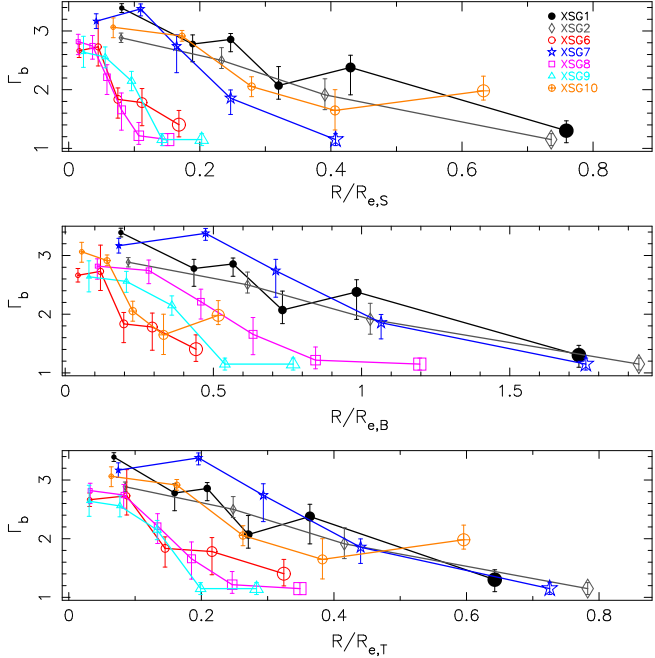}
 \end{center}
    \caption{Same as Fig.~\ref{fig:IMF_ref} but rescaling the galactocentric distance, $\rm R$,
    by the  galaxy effective radius. We consider (top) effective radii from 
    single S\'ersic fits of galaxy images, (middle) effective radii of the bulge component only (from B+D decomposition), and (bottom) ``total'' effective radii from B+D models (see Sec.~\ref{sec:spmodels} for details). Effective radii for different galaxies are reported in Tab.~\ref{tab:strucpars}. Notice the different x-axis scale of the three panels, due to differences 
    in $\rm R_e$ among different methods.
        }
    \label{fig:imf_slope_radii}
\end{figure}

As reported in Sec.~\ref{subsec:IMFgradients},  while IMF profiles of the XSGs appear to be all similar when plotted versus galactocentric distance in physical units, the profiles do differ significantly when normalizing distances by the effective radius, $\rm R_e$. This is shown in Fig.~\ref{fig:imf_slope_radii}, where we plot the best-fitting IMF slope, for our reference fitting method, versus $\rm R/R_e$, considering effective radii from single S\'ersic fits (top panel), for the bulge component only (middle panel), and from B+D models. As discussed in Sec.~\ref{sec:spmodels}, most X-Shooter galaxies exhibit an extended outer envelope making the estimate of $\rm R_e$ strongly dependent on the adopted method. In particular, XSG6, XSG8, and XSG9 have single S\'ersic and B+D effective radii significantly larger than those of the other galaxies. Hence, their IMF profiles, when plotted as a function of $\rm R/R_e$, appear more concentrated than those of XSG1, XSG2, XSG7, and XSG10 (see top and bottom panels in Fig.~\ref{fig:imf_slope_radii}). This implies, in turn, that within a given fraction of $\rm R_e$, the (luminosity-weighted) IMF slope and mass-to-light ratio of  the XSG's would show a large scatter.

\section{Best-fitting and observed line-strengths.}
\label{app:bestfitindices}

\begin{figure*}
 \begin{center}
\leavevmode
    \includegraphics[width=17.5cm]{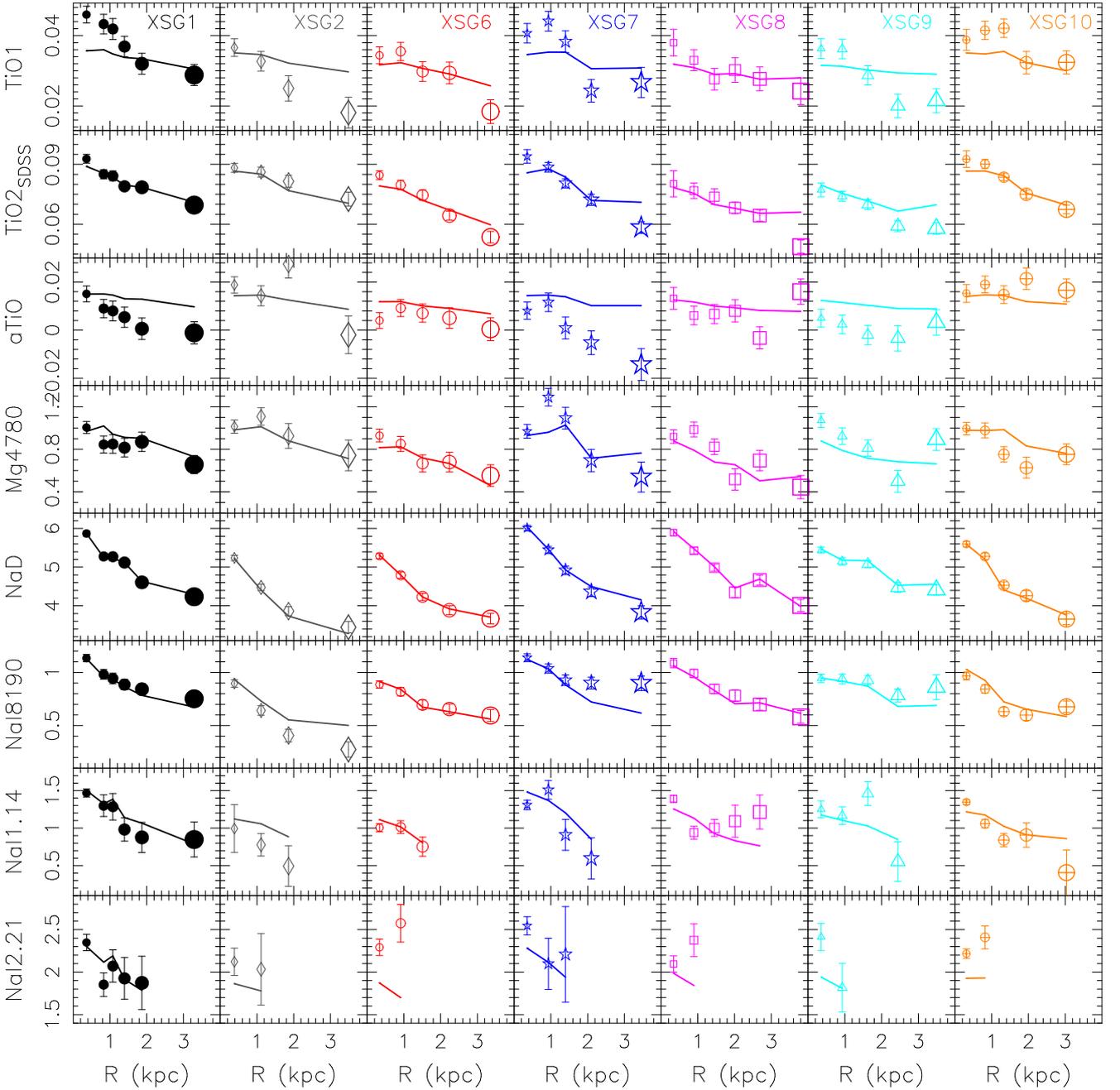}
 \end{center}
    \caption{Observed and best-fitting line-strengths as a function of galactocentric distance, for our reference fitting scheme (method A; see Tab.~\ref{tab:methods}). Observed line-strengths are plotted as symbols with error bars, while best-fitting indices are given by the solid lines. Only IMF-sensitive features are shown in the plot. Different galaxies are plotted from left to right (see labels in the top panels), while different indices are plotted along each column in the Figure, from top (\tioi) to bottom (\naiv ). Line-strengths of different galaxies are plotted with different symbols (whose size increases with galactocentric distance) and colours, as in Fig.~\ref{fig:IMF_ref}. Error bars denote 1$\sigma$ uncertainties.
        }
    \label{fig:fit_indices}
\end{figure*}

\begin{figure*}
 \begin{center}
\leavevmode
    \includegraphics[width=17.5cm]{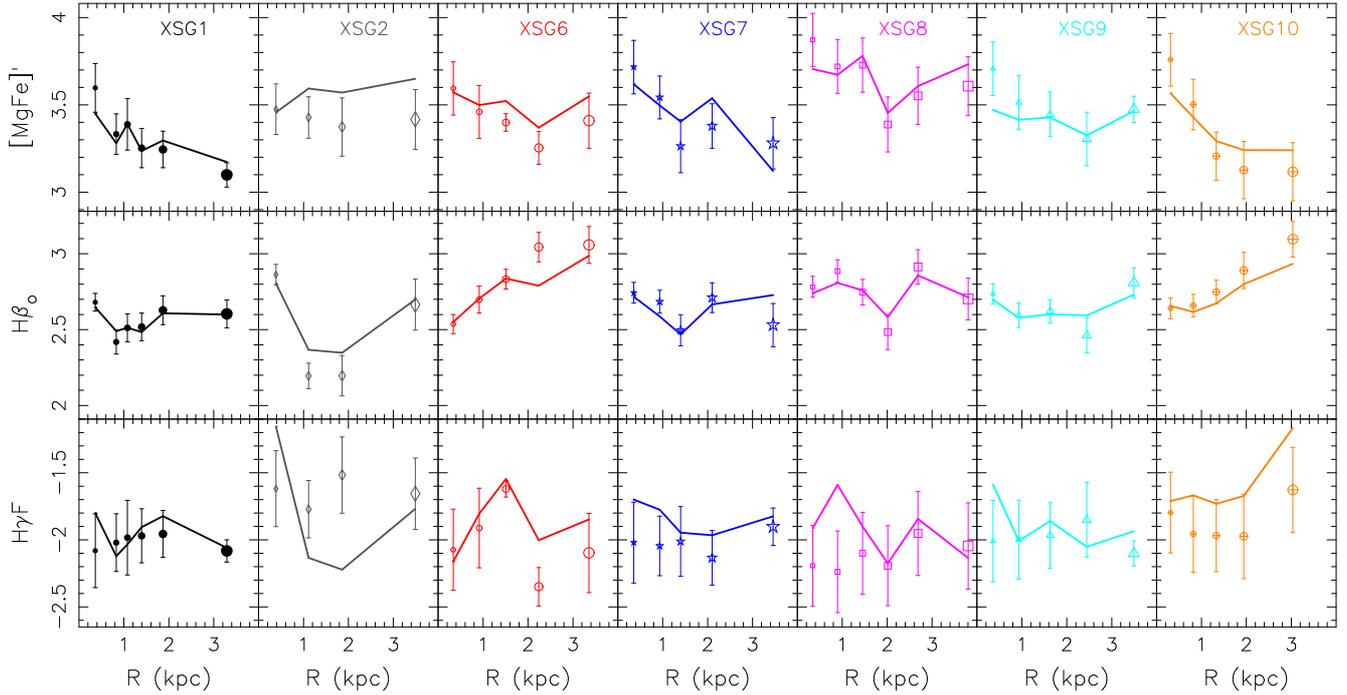}
 \end{center}
    \caption{Same as Fig.~\ref{fig:fit_indices} but for the total metallicity indicator \mgfep (top row), and \hbo\ and \hgf\ Balmer lines (middle and bottom rows, respectively). Notice that for \mgfep\ we plot best-fitting line-strengths from our reference fitting scheme (as in Fig.~\ref{fig:fit_indices}), while for the Balmer lines we consider results for method D (see Tab.~\ref{tab:methods}), as Balmer lines are not included in the fitting for the other methods.
        }
    \label{fig:fit_indices_mgfep_balmer}
\end{figure*}

Fig.~\ref{fig:fit_indices} compares observed and best-fitting IMF-sensitive spectral indices of all X-Shooter spectra, for our reference fitting method. In Fig.~\ref{fig:fit_indices_mgfep_balmer}, we also show fitting results for the total-metallicity indicator \mgfep, as well as for \hbo\ and \hgf\ Balmer lines. Notice that all best-fitting  indices are corrected to a broadening of 300~\kms\ velocity dispersion, following a similar procedure as in LB16 and LB17 (i.e. using best-fitting stellar population models to evaluate the a broadening correction to line-strengths). Results are shown for our reference fitting method (method A), but for \hbo\ and \hgf , as Balmer lines are only included in the 2SSP fitting procedure. Hence, for \hbo\ and \hgf, Fig.~\ref{fig:fit_indices_mgfep_balmer} overplots best-fitting indices from method D (see Tab.~\ref{tab:methods}).

Fig.~\ref{fig:fit_indices}  shows that we fit reasonably well all spectral indices, although some discrepancies between models and observations are actually found: 
\begin{description}
\item[--] TiO1 shows a steeper trend in the data, with respect to  models. This might be due to (i) the effect of non-solar abundance ratios, which is not captured by our modeling approach; (ii) the fact that TiO1 is a very broad feature, whose radial gradient might be affected by sky-continuum subtraction; (iii) the extrapolation of the feature in the high metallicity regime (see also LB16);
\item[--] aTiO shows some offset/tilt with respect to observations (in particular for XSG1, XSG7, and XSG9). Notice that aTiO is a very broad feature, even broader than the TiO's. Hence sky-continuum subtraction/flux calibration uncertainties (at one percent level) can significantly affect its radial behaviour. Moreover, as discussed in LB16, the competing effect of IMF and \zh\ does not make this feature an IMF-indicator as good as other features. 
\item[--] Remarkably, we are able to fit all four Na lines, simultaneously, at different radial positions. This confirms the finding of LB17, that the combined effect of a bottom-heavy IMF and overabundant \nafe\ is crucial to describe Na features. Nevertheless, we notice that for most galaxies, \naiv\ is underpredicted by our models. This is somewhat expected, as theoretical stellar spectra used to construct Na--EMILES models do not cover temperatures cooler
than \Teff=3500\,K (see LB17). Hence, for stars with \Teff$<$3500\,K in the empirical stellar libraries (see Sec.~\ref{sec:spmodels}), we apply differential corrections for
\nafe\ by assuming \Teff=3500\,K. At the SSP level, this implies that NIR Na features (and in particular \naiv ) might be underestimated by $\sim 0.1$--$0.2$~dex (see LB17), which is fully consistent with what seen in the bottom panels of Fig.~\ref{fig:fit_indices}. Moreover, since \cfe\ is over-abundant in massive ETGs, the underestimation of \naiv\ might also due to a significant contribution of carbon to \naiv, as discussed in~\citet{Benny:2017}.
\end{description}

We also notice that Balmer lines are well matched by our 2SSP fitting scheme, implying that our IMF results are also insensitive to age constraints. Indeed, the only galaxies with some evidence of age gradients are XSG6 and XSG10, where \hbo\ increases significantly with galacto-centric distance. The presence of a (negative) age gradient for XSG6 is also confirmed by our reference fitting method A (see the age of XSG6 decreasing outwards in panel e of Fig.~\ref{fig:gammab_gradients}), where none of the Balmer lines is included in the fitting procedure, as well as by the presence of significant emission lines in this galaxy at all radii (see Appendix~\ref{app:emission}). For XSG10, the situation is less clear, as this galaxy does not show any significant emission (see Appendix~\ref{app:emission}), and method A does not provide any significant age gradient (see panel e of Fig.~\ref{fig:gammab_gradients}). Therefore, for XSG10, the \hb\ radial gradient might be  reflecting an IMF, rather than an age, variation (in fact, \hbo\ becomes weaker with increasing IMF slope; see LB13). Whatever is causing the radial trend of \hbo\ for XSG10, the key point for the present work is that for this galaxy, methods A and  D provide very consistent IMF radial variations. More in general, the fact that different combinations of spectral indices (Tab.~\ref{tab:methods}) give similar IMF radial trends for all galaxies, proves the robustness of our results, despite some unavoidable uncertainty in (state-of-art) stellar population models.

\section{Emission lines in the X-Shooter spectra.}
\label{app:emission}
Some X-Shooter spectra of galaxies in our sample show emission contamination in the Balmer lines, and/or other weak emission lines in the optical spectral range. In order to characterize the nature of the emission,  we have used the BPT~\citep{BPT} and WHAN (see~\citealt{Cid:2004} and references therein) diagnostic diagrams, fitting the relevant emission lines, i.e. H$\beta$, [OIII]$\lambda$5007, [NII]$\lambda$6548, H$\alpha$, and [NII]$\lambda$6584; the latter, though not explicitly used in the BPT/WHAN diagrams, is analyzed as it can affect the H$\alpha$ line due to the small relative separation of [NII]$\lambda$6584 and H$\alpha$. The fits are performed with a dedicated Python code, after subtracting the best-fitting model of the stellar component to each spectrum, within spectral windows large enough to ensure that both the emission line profiles and a significant portion of the continuum are included in the fitting range. The stellar component is modeled with a linear superposition of MILES SSP models, with varying age and metallicity, combined with CvD12a response functions to account for the effect of non-solar abundance ratios. The fits include a constant tilt in the residual continuum around the emission lines, and are performed by treating either (a) velocity ,$\rm v_{em}$, and velocity dispersion, $\rm \sigma_{em}$, as independent fitting parameters for each emission line; or (b) fixing  $\rm v_{em}$ and $\rm \sigma_{em}$ to be the same for all lines. We have verified that all the results reported here are independent of the assumptions on $\rm v_{em}$ and $\rm \sigma_{em}$.

\begin{figure}
\includegraphics[width=8cm]{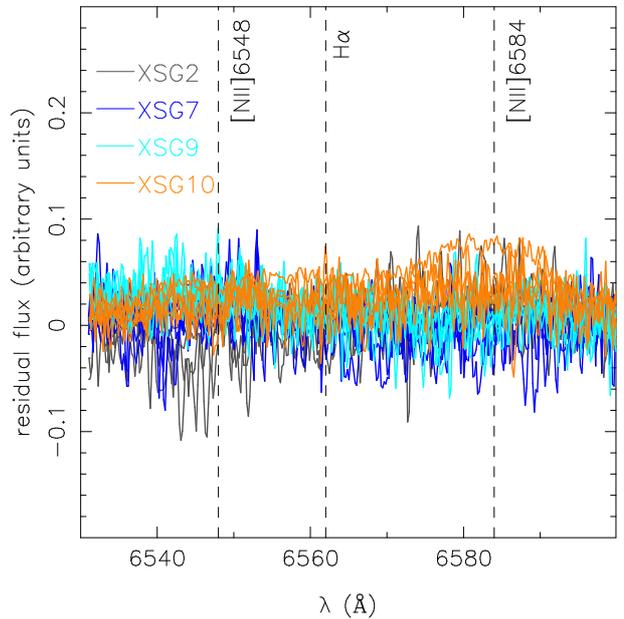}
\caption{Residual spectra (after subtracting the best-fit stellar spectrum) for all radial bins of XSG2, XSG7, XSG9, and XSG10, in the restframe spectral window around the H$\alpha$ line, from 6530 to 6600\AA. The position of the emission lines [NII]$\lambda$6548, H$\alpha$, and [NII]$\lambda$6584 are marked with dashed vertical lines. No emission pattern is detected for these galaxies. Different galaxies are plotted with the same colour coding as in Fig.~\ref{fig:IMF_ref}.}
\label{fig:noemission_gals}
\end{figure}

\begin{figure}
\includegraphics[width=8cm]{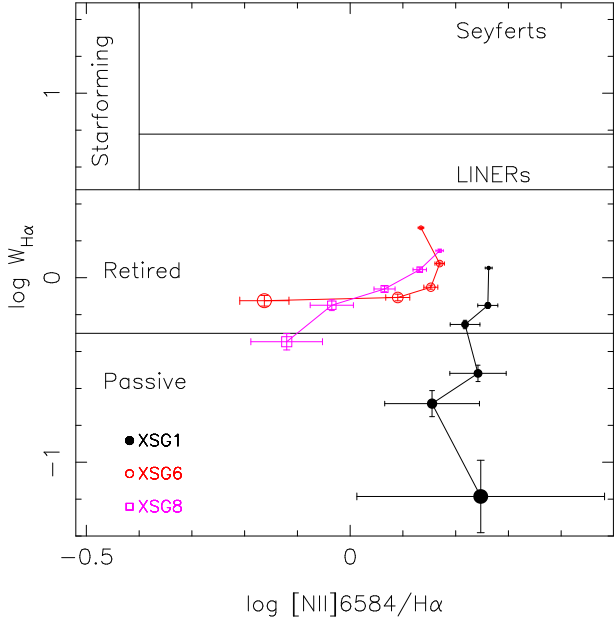}
\caption{WHAN diagram for XSG1, XSG6, and XSG8. Each galaxy is plotted with the same symbol types and colours as in Fig.~\ref{fig:IMF_ref}. Symbol sizes increase with galactocentric distance.  For these galaxies, the ionization pattern tends to vary from retired   in the center, to passive in the outskirts. For the other XSGs (XSG2, XSG7, XSG9, and XSG10), no emission is detectable in our spectra.}
\label{fig:whan}
\end{figure}

Most galaxies in our sample (XSG2, XSG7, XSG9 and XSG10) are found to show negligible emission, as illustrated in Fig.~\ref{fig:noemission_gals}, where we plot the residual flux (after subtracting the stellar component)  in the spectral regions around the H$\alpha$ line, for all radial bins. On the other hand, XSG1, XSG6, and XSG8, present detectable emission lines, whose ionization pattern can be characterized through the WHAN diagram, as shown in Fig.~\ref{fig:whan}:
\begin{itemize}
 \item[--]  XSG1 shows emission lines up to the largest radial bin probed, also detectable in the [OII]$\lambda$3726-3728 and [SII]$\lambda$6716-6730 lines. The WHAN diagram shows that the galaxy is retired (emission line produced by hot post-asymptotic giant branch stars) in the center and passive~\footnote{Notice that, as discussed in~\citet{Cid:2004}, the separation between retired and passive galaxies is somewhat arbitrary, as both classes include similar objects. ``Passive'' galaxies are those objects for which emission lines are extremely weak, making their measurement less safe.} in the outer regions
 (see filled circles in Fig.~\ref{fig:whan});
 \item[--] XSG6 also has intense [OII]$\lambda$3726-3728, [SII]$\lambda$6716-6730 and [OI]$\lambda$6300 emission.  The WHAN diagram points to a retired ionization pattern for all apertures (see open circles in the Figure);
 \item[--] As XSG1, XSG8 also presents detectable [OII]$\lambda$3726-3728 and [SII]$\lambda$6716-6730 emission, at least in the most central radial bins. The WHAN diagram indicates shows a retired ionization
pattern, but for the last measured aperture where the system is passive (see open squares in the Figure).
\end{itemize}
Although we have only a few galaxies with emission, we notice that XSG1 has a more concentrated gas distribution compared to XSG6 and XSG8 where from inner to outer regions emission is always in the ``retired'' region of the WHAN diagram. Another way of interpreting is that since XSG6 and XSG8 also have much larger $\rm R_e$ than XSG1, this trend may just be reflecting how gas traces stars. 

We verified that all results reported above from the WHAN diagram are consistent with those from the BPT diagram, that is not shown here for brevity reasons. 
Hence, to summarize,  XSG1, XSG6, and XSG8 show a ionization pattern that varies  from retired  to passive with increasing galactocentric distance, while the other XSGs do not show any measurable emission. Notice that, as discussed in the text, no significant dependence of IMF radial variations are found between galaxies with different ionization patterns.

\bsp	
\label{lastpage}
\end{document}